\documentclass[12pt]{article}
\usepackage{latexsym}
\usepackage{epsfig}
\csname @addtoreset\endcsname{equation}{section}

\def\bseq{\begin{subequation}}  
\def\eseq{\end{subequation}}
\def\bsea{\begin{subeqnarray}}  
\def\esea{\end{subeqnarray}}

\newcommand{\1}{ \,  \raisebox{+0.14em}{{\hbox{{\rm \scriptsize ]}} \raisebox{-0.2em}{\kern-.8em\hbox{1}}}} \, }

\newcommand{\bbox}{\lower.2ex\hbox{$\Box$}}
\newcommand{\eqn}[1]{(\ref{#1})}

\newcommand{\bde}{\begin{description}}
\newcommand{\ede}{\end{description}}

\renewcommand{\a}{\alpha}
\renewcommand{\b}{\beta}

\renewcommand{\d}{\delta}

\newcommand{\pa}{\partial}
\newcommand{\g}{\gamma}
\newcommand{\G}{\Gamma}

\newcommand{\e}{\epsilon}

\newcommand{\m}{\mu}

\newcommand{\n}{\nu}

\newcommand{\s}{\sigma}

\renewcommand{\t}{\tau}

\renewcommand{\sb}{\bar{\sigma}}

\newcommand{\Db}{\bar{D}}
\newcommand{\Wb}{\bar{W}}

\newcommand{\fb}{\bar{f}}

\newcommand{\ad}{{\dot{\alpha}}}
\newcommand{\bd}{{\dot{\beta}}}
\newcommand{\gd}{{\dot{\gamma}}}
\newcommand{\dd}{{\dot{\delta}}}

\newcommand{\DC}{\nabla}

\newcommand{\Tr}{{\rm Tr}}

\newcommand{\Tra}{{\rm{Tr}}_{Ad}}

\newcommand{\wt}[1]{\widetilde{#1}}

\begin{document}
\begin{titlepage}
\begin{flushright}
IFUM-FT-688\\

\end{flushright}
\vspace{2cm}
\noindent{\Large \bf $F^5 $ contributions to the nonabelian Born Infeld action }\\
\vskip 0.2mm
\noindent{\Large \bf from a supersymmetric Yang-Mills
five-point function}
\vspace{1cm}
{\bf \hrule width 16.cm}
\vspace {1cm}

\noindent{\large \bf Andrea Refolli, Alberto Santambrogio, Niccolo' Terzi, \\
Daniela Zanon}

\vskip 2mm
{ \small

\noindent Dipartimento di Fisica dell'Universit\`a di Milano
and

\noindent INFN, Sezione di Milano, Via Celoria 16,
20133 Milano, Italy}
\vfill
\begin{center}
{\bf Abstract}
\end{center}
{\small We consider the ${\cal N}=4$ supersymmetric Yang-Mills theory in four dimensions. We compute the one-loop contributions to the effective action with five external vector fields and compare them with corresponding results in open superstring theory. Our calculation determines the structure of the $F^5 $ terms that appear in the nonabelian generalization of the Born Infeld action.  The trace operation on the gauge group indices receives contributions from the symmetric as well as the antisymmetric part. 
We find that in order to study corrections to the symmetrized trace prescription one has to consistently take into account derivative contributions not only with antisymmetrized products $\nabla_{[\m }\nabla_{\n]}$ but also with symmetrized ones $\nabla_{(\m }\nabla_{\n)}$. }
\vspace{2mm}
\vfill \hrule width 6.cm
\begin{flushleft}
e-mail: andrea.refolli@mi.infn.it\\
e-mail: alberto.santambrogio@mi.infn.it\\
e-mail: niccolo.terzi@mi.infn.it\\
e-mail: daniela.zanon@mi.infn.it
\end{flushleft}
\end{titlepage}

\section{Introduction}
Scattering amplitudes of massless modes in string theories can be described in terms of effective  Lagrangians. For the open string case the abelian Born-Infeld action \cite{BI} represents a remarkable example of an effective field theory which contains string corrections to all orders in $\a'$ \cite{FT,ACNY,Tseytlin3}. The contributions can be summed up to all orders
in the case of a constant, abelian field strength. As soon as these two 
conditions are relaxed the problem becomes complicated and a complete action for such fields has not been obtained. A non-abelian generalization of the Born-Infeld action can be defined as suggested in \cite{Tseytlin1} using a symmetrized trace operation over the gauge group matrices. However there are indications that this prescription might not be sufficient to include all the contributions that one would obtain from an open string approach \cite{HT,BRS,STT}.

The best available way to  construct the effective action for the non-abelian, non constant curvature case, is to proceed order by order. One has to compute corrections with increasing number of derivatives and of field strengths. A well established result is up to the order $\a'^2$ \cite{GW,Tseytlin2}.
As soon as one focuses on higher orders the calculations become difficult. The inclusion of supersymmetry seems to be quite useful to gain insights and it might set enough constraints to fix uniquely the form of the action. Several attempts in different directions are under consideration \cite{BRS,STT,molti}.

In this paper we attack the problem as follows: \\
since we want to make contact with the ten-dimensional open superstring we consider its four-dimensional field theory limit, i.e. the ${\cal N}=4$ supersymmetric Yang-Mills theory. Then we compute at one-loop perturbatively: the idea is that in this way we construct an effective action which is supersymmetric and generalizes the Yang-Mills theory. If supersymmetry determines the form of the allowed deformations \cite{uniqueness} it should correspond to
 the non-abelian Born-Infeld theory. We set the external background on-shell and
compute one-loop $n$-point vector amplitudes. These functions are in general non-local expressions that contain a loop-momentum integral. For a given $n$-point amplitude we perform a low-energy expansion in the external momenta: this is achieved through the introduction of an infrared mass regulator that in the expansion plays the same role as $\a'$, i.e. it keeps track of the order of the derivative term under consideration. The net result is that each $n$-point amplitude gives rise to an infinite series of local terms with higher  and higher number of derivatives. The leading term is automatically gauge invariant. The subleading  contributions which contain derivatives are not gauge-invariant by themselves. The correct covariantization is obtained including corrections from amplitudes with a higher number of external background fields through a mechanism similar to the one in \cite{PSZ} .

We present the calculation of the four- and five-point functions, and of the part of the six-point function which is needed for the above mentioned covariantization of the lower order results. These computations will allow us to evaluate the complete gauge invariant structures for the  $F^4$ and the $\nabla\nabla F^4$, $F^5$ contributions. Since the symmetric trace prescription would rule out $F^5$
terms, the nonvanishing result that we find confirms that the actual form of the non-abelian Born-Infeld action is definitely richer. In fact the trace operation on the gauge group indices receives contributions from the symmetric as well as the antisymmetric part. 

In the next section we briefly recall the ${\cal N}=1$ superspace formulation of the ${\cal N}=4$  Yang-Mills action and describe the main ingredients that enter the quantization via the background field approach. Moreover we outline the general procedure that, starting  from the one-loop Yang-Mills amplitudes, allows us to determine corrections to the nonabelian Born-Infeld action.
In section 3 we compute the one-loop amplitude with four external vector fields and study its low-energy expansion. We extract from the superfield result its component content and in particular we study the bosonic contributions which contain the field strengths $F_{\m\n}$. In section 4 we compute in the same manner the five-point function. The calculation is quite complicated, on one hand because of the gauge group structure, on the other hand because  of the existence of several on-shell identities which  makes it difficult to express the result in a canonical form. Finally in section 5 we compute the part of the six-point function that we use to complete the covariantization of the terms with four field strengths and two derivatives.
In section 6 we collect all our results and  compare them with corresponding results from open superstring theory \cite{kitazawa}.

\section{The ${\cal N}=4$  Yang-Mills action and its quantization}

The ${\cal N}=4$ supersymmetric Yang-Mills  classical action
written in terms of ${\cal N}=1$ superfields (we use the
notations and conventions adopted in \cite{superspace}) is given by
\begin{eqnarray}
&&S= \frac{1}{g^2}~{\rm Tr} \left( \int~ d^4x~d^4\theta~ e^{-V}
\bar{\Phi}_i e^{V} \Phi^i +\frac{1}{2} \int ~d^4x~d^2\theta~ W^\a W_\a
+\frac{1}{2} \int ~d^4x~d^2\bar{\theta}~ \bar{W}^\ad \bar{W}_\ad\right.\nonumber\\
&&\left.~~~~~~~~~~~~~+\frac{1}{3!} \int ~d^4x~d^2\theta~ i\e_{ijk}
 \Phi^i
[\Phi^j,\Phi^k] + \frac{1}{3!}\int ~d^4x~d^2\bar{\theta}~ i\e^{ijk} \bar{\Phi}_i
[\bar{\Phi}_j,\bar{\Phi}_k] \right)
\label{N4SYMaction}
\end{eqnarray}
where the $\Phi^i$ with $i=1,2,3$ are three chiral
superfields, and the $W^\a= i\bar{D}^2(e^{-V}D^\a e^V)$ are the gauge
superfield strengths. All the fields are Lie-algebra valued, e.g.
$\Phi^i=\Phi^i_a T^a$, in the adjoint representation of the gauge group, with $[T_a,T_b] = i f_{abc} T_c$. 

At the component level the three chiral superfields $\Phi^i$ contain  six spin-$0$ particles and three spin-$\frac{1}{2}$ Weyl spinors,
while $W^\a$ describes one spin-$\frac{1}{2}$ spinor and one gauge vector particle.
Since we are interested in computing field strength corrections to the Born-Infeld action, we will consider amplitudes with vector fields as external background. Moreover we will extract from them only the gauge field bosonic components using the relations 
\begin{eqnarray}
&&D_{(\a}W_{\b)}|_{\theta=0}=f_{\a\b}=\frac{1}{2} (\sigma_{\m\n})_{\a\b}F^{\m\n}\nonumber\\
&&\Db_{(\ad}\Wb_{\bd)}|_{\theta=0}=\bar{f}_{\ad\bd}=-\frac{1}{2} (\bar{\sigma}_{\m\n})_{\ad\bd}F^{\m\n}
\label{component}
\end{eqnarray}
with $\sigma_{\m\n}$ and $\bar{\sigma}_{\m\n}$ defined as in (\ref{sigmamatr}). 
In this way we will be able to isolate the relevant contributions. 

In order to evaluate one-loop amplitudes we need to quantize the theory and this is most efficiently done using  the background field method
\cite{GS,GZ} which guarantees a gauge invariant result for the effective action.
For the gauge multiplet one performs a non linear splitting between
the quantum prepotential $V$ and the background superfield, via covariant
derivatives (in quantum chiral representation \cite{superspace})
\begin{eqnarray}
&&\nabla_\a =e^{-V} D_\a~ e^V
 ~\rightarrow~e^{-V}\nabla^B_\a ~ e^V
\nonumber\\
&&~~~~\nonumber\\
&& \bar{\nabla}_\ad= \bar{D}_\ad
~\rightarrow~
 \bar{\nabla}^B_\ad
\label{backcovder}
\end{eqnarray}
In this way the external background enters the quantum action implicitly
in the background covariant derivatives through the connections
\begin{equation}
\nabla^B_\a= D_\a-i{\bf{\G}}_\a \qquad \qquad \bar{\nabla}^B_\ad= \bar{D}_\ad-
i\bar{\bf{\G}}_\ad \qquad\qquad \nabla^B_a=\pa_a-i{\bf{\G}}_a
\label{backcovderconn}
\end{equation}
and explicitly in the background field strength
${\bf W}_\a=\frac{i}{2}[\bar{\nabla}^{B\ad},\{\bar{\nabla}^B_\ad,\nabla^B_\a\}]$.
Background covariant gauge-fixing introduces additional terms
\begin{equation}
-\frac{1}{2g^2} {\rm Tr} \int
d^4x~d^4\theta~V(\nabla_B^2\bar{\nabla}_B^2+\bar{\nabla}_B^2\nabla_B^2)V
+S_{FP}+S_{NK}
\label{gauge-fixing}
\end{equation}
with Faddeev-Popov action
\begin{equation}
S_{FP}= {\rm Tr}\int d^4x~d^4\theta~[\bar{c}'c-c'\bar{c}+\frac{1}{2}
(c'+\bar{c}')[V,c+\bar{c}]+\dots]
\label{FP}
\end{equation}
and Nielsen-Kallosh ghost action
\begin{equation}
S_{NK}={\rm Tr}\int d^4x~d^4\theta~\bar{b}{b}
\label{NK}
\end{equation}
The three ghosts $c$, $c'$ and $b$ are background covariantly chiral
superfields, i.e. $\bar{\nabla}_B^\ad c=\bar{\nabla}_B^\ad c'
=\bar{\nabla}_B^\ad b=0$. In the following we drop the suffix $B$ from the 
covariant
derivatives. 

After gauge-fixing the quadratic quantum $V$-action becomes
\begin{eqnarray}
&&S\rightarrow -\frac{1}{2g^2} {\rm Tr} \int
d^4x~d^4\theta~V\left[\Box-i {\bf \G}^{\g\gd}\pa_{\g\gd} -\frac{i}{2}
(\pa^{\g\gd}{\bf \G}_{\g\gd})-\frac{1}{2}{\bf \G}^{\g\gd}{\bf \G}_{\g\gd}\right.
\nonumber\\
&&~~~~~~~~~~~~~~~~~~~~~~~~~~~~~~~~~\left.-i {\bf W}^\a(D_\a-i{\bf\G}_\a)-i\bar{\bf W}^\ad
(\bar{D}_\ad-i{\bf\G}_\ad)\right]V
\label{actionquadratic}
\end{eqnarray}
As mentioned above we want to compute one-loop contributions to the effective action with external vector fields. The ${\cal N}=4$ theory is particularly simple since this type of terms are produced only by quantum vector loops. The loops with the three chiral matter fields are cancelled by the three ghosts: in fact each ghost
contributes to the
one-loop effective action exactly as
a standard chiral superfield would do, the only
difference being an overall opposite sign because of the statistics.

Thus we focus on the action in (\ref{actionquadratic}). The
interactions with the background
are at most linear in the $D$-spinor derivatives and at least two
$D$'s and two $\bar{D}$'s are needed in the loop in order to
complete the $D$-algebra. Thus 
the first non-vanishing result is at the level of the four-point function \cite{GS2}.
Here we describe the general procedure we have adopted in order to extract from
the super Yang-Mills effective action the wanted contributions to the non-abelian  Born-Infeld action.

A one-loop $n$-point amplitude will contain $n$ external background fields from the interactions in (\ref{actionquadratic}), with at least two ${\bf W}$ and two $\bar{\bf W}$ in order to complete the $D$-algebra.
We label the external momenta $p_1,p_2,\dots,p_n$
and introduce a general notation for the dependence on the gauge group (with structure constants $f_{abc}$)
\begin{eqnarray}
g(a_1,a_2,\ldots,a_n)=f_{x_1 a_1 x_2}f_{x_2 a_2 x_3}\cdots f_{x_n a_n x_1}
\label{colourstructure}
\end{eqnarray}  
In the following we treat the gauge fields as matrices in the adjoint representation, i.e. ${\bf{W}}^\a_{ac}\equiv f_{abc} {\bf{W}}^{\a b}$. We define
\begin{equation}
\Tra(AB\cdots )\equiv A_a B_b \cdots  g(a,b,\ldots)  
\label{trace}
\end{equation} 
Moreover we set the external fields on-shell, i.e. we freely use the equations of motion
\begin{equation}
\nabla^\a {\bf W}_\a=0\qquad\qquad\qquad  \bar{\nabla}^\ad {\bf \Wb}_\ad=0  
\label{equationmotion}
\end{equation}                
In this way a $n$-point function gives rise to a background field dependence which contains a total of $n$ fields, with a given number of connections ${\bf \G}$, and of field strengths ${\bf W}$, ${\bf \Wb}$, of the general form
\begin{equation}
\int d^4\theta~ \Tra({\bf W}^{\a}(p_1)\dots{\bf \Wb}^{\ad}(p_i)\dots{\bf W}_{\a}(p_j)\dots{\bf \G}^{\g\gd}(p_k)\dots{\bf \Wb}_{\ad}(p_l)\dots \bar{\nabla}_{\bd}{\bf \Wb}_{\dot{\d}}(p_n)) 
\label{backdep}
\end{equation}
From (\ref{backdep}) we want to extract the relevant structures that might produce $F_{\m\n}$'s at the component level. To this end we need convert 
\begin{equation}
\int d^4\theta\equiv\int d^2\theta~d^2\bar{\theta}\Rightarrow \frac{1}{4} D^\a D_\a \Db^\ad \Db_\ad
\label{thetaint}
\end{equation}
and act with these four spinor derivatives on the ${\bf W}$ and ${\bf \Wb}$ using 
the relations in (\ref{component}). One can reconstruct gauge covariant expressions also from the connections ${\bf \G}$, but this can happen only at the level of the six-point function and beyond. Since in this paper we consider structures up to the five-point functions we leave this point for a future discussion.

In conclusion we obtain the structures we are looking for if our one-loop diagram has produced products of fields $D_{(\a}{\bf W}_{\b)}$ and their hermitian conjugate ones $\Db_{(\ad}{\bf \Wb}_{\bd)}$. In addition we have to deal with a loop-momentum integral which contains $n$ scalar propagators and momentum factors directly from the vertices in (\ref{actionquadratic}) and/or from commutators of spinor derivatives produced while performing the $D$-algebra
\begin{equation}
I_n=\int {\rm d}^4 k \frac{h_n(k,p_i)}{k^2 (k+p_2)^2(k+p_2+p_3)^2\dots(k+p_2+\dots+p_n)^2}
\label{npointint}
\end{equation}
We can rewrite (\ref{npointint}) as an infinite series of local terms in a low-energy expansion with higher derivatives. We introduce an IR mass $M$ and expand the propagators keeping the external momenta small as compared to $M$. First we Feynman combine the propagators in (\ref{npointint})
\begin{eqnarray}
  \label{eq:feynmann}
\frac{1}{A_1 A_2 \cdots A_n} & = & \Gamma(n) \int_0^1 {\rm d}x_1 \int_0^{x_1}{\rm d}x_2 \cdots \int_0^{x_{n-2}} {\rm d}x_{n-1} \nonumber \\
&& \left[A_1 x_{n-1} + A_2 (x_{n-2}-x_{n-1}) + \cdots + A_n (1-x_1) \right]^{-n}
\end{eqnarray}
Then shifting the $k$-momentum we obtain
\begin{eqnarray}
&&\frac{1}{k^2 (k+p_2)^2(k+p_2+p_3)^2\dots(k+p_2+\dots+p_n)^2}\nonumber\\
&& \nonumber \\
&&~~~~~~~~~~ \Rightarrow (n-1)!\int {\rm d}^4 k  \int_0^1 {\rm d}x_1 \int_0^{x_1} {\rm d}x_2 \dots \int_0^{x_{n-2}} {\rm d}x_{n-1}  \nonumber \\
&&~~~~~~~~~~~~~~~~~\left[
k^2 + M^2 +(1-x_1)(p_2+\dots+p_n)^2+\dots+(x_{n-2}-x_{n-1})(p_2)^2\right. \nonumber \\
&& ~~~~~~~~~~~~~~~~~~~~~~~~~~~~~~~~~~~~\left.-\left((1-x_{n-1})p_2
+\dots+(1-x_1)p_n\right)^2\right]^{-n} \nonumber \\
&& \nonumber \\
&& ~~~~~~~~~~=\frac{1}{(k^2 + M^2)^n} -n! \int {\rm d}^4 k \int {\rm d}x_i \frac{H_n(x_i,p_j)}{(k^2+M^2)^{n+1}} + \ldots
\label{nlenergyexpansion}
\end{eqnarray}
where
\begin{eqnarray}
H_n(x_i,p_j)&=&(1-x_1)(p_2+\dots+p_n)^2+\dots+(x_{n-2}-x_{n-1})(p_2)^2\nonumber \\
&& ~~~~~~~~~~~~-\left[(1-x_{n-1})p_2+\dots+(1-x_1)p_n\right]^2
\label{Hn}
\end{eqnarray}
All the integrals over the $k$-momentum are finite and give
\begin{eqnarray}
&&\int d^4k \frac{1}{(k^2 + M^2)^n}= \frac{\pi^2}{(n-2)(n-1)}\frac{1}{(M^2)^{n-2}}\nonumber\\
&&~~~\label{kintegrals}\\
&&\int d^4k \frac{k_{\a\ad}k_{\b\bd}}{(k^2 + M^2)^n}=\frac{\pi^2}{(n-3)(n-2)(n-1)}\frac{1}{(M^2)^{n-3}} C_{\a\b}C_{\ad\bd} \nonumber\\
&&~~~~\nonumber\\
&&\dots\dots\dots \nonumber
\end{eqnarray}
The non-local expression $I_n$
 has been traded by an infinite sum of local terms.
In the next two sections we apply this procedure to the computation of the four- and the five-point functions and show explicitly how we can determine order by order structures that we interpret as corrections to the non-abelian Born-Infeld action.
\section{The four-point function}
Now we consider one-loop contributions with four external vector fields.
They correspond to box-type diagrams as the one shown in Fig.1. We label the external momenta  $p_1,p_2,p_3,p_4$
and use the notation introduced in (\ref{colourstructure})  
with $n=4$. 
A diagram like the one in Fig.1 gives rise to a momentum integral
\begin{eqnarray}
I_4 &=& \int {\rm d}^4 k \frac{1}{k^2 (k+p_2)^2(k+p_2+p_3)^2(k+p_2+p_3+p_4)^2} 
\label{box}
\end{eqnarray}
and to a background field dependence
\begin{equation}
\int d^4\theta ~\Tra({\bf W}^{\a}(p_1){\bf W}_{\a}(p_2){\bf \Wb}^{\ad}(p_3){\bf \Wb}_{\ad}(p_4)) 
\label{4background}
\end{equation}
The complete answer is obtained by summing to the above expression all permutations in the $2,3,4$ indices. 
\begin{figure}[htb] \begin{center}
\mbox{\epsfig{figure={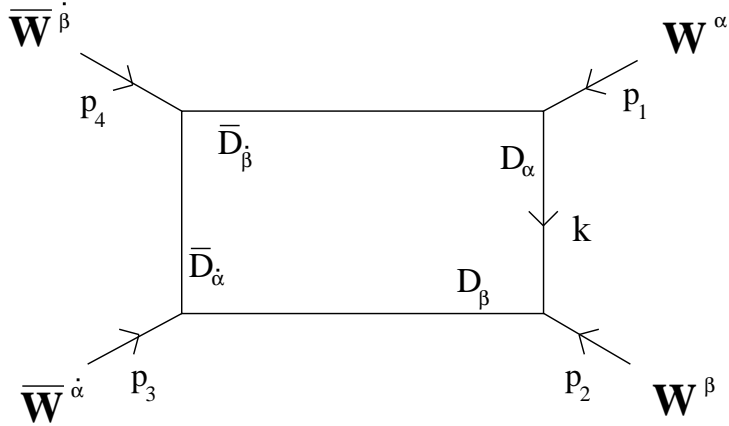},
       width=0.6 \linewidth}}
\caption{Four-point amplitude}
\label{Box} \end{center}
\end{figure}
We are interested in rewriting the superspace result in components and in particular we want to study the bosonic gauge field contributions. To this end we use the definitions in (\ref{component}) and the relations in the Appendices. First we obtain
\begin{eqnarray}
&&\int d^4\theta ~{\bf W}^{\a}(p_1,a_1){\bf W}_{\a}(p_2,a_2){\bf \Wb}^{\ad}(p_3,a_3){\bf \Wb}_{\ad}(p_4,a_4) \nonumber\\
&&~~~~~ \Rightarrow  (f_1)_{\a}^{~\b}(f_2)_{\b}^{~\a}(\fb_3)^{\ad}_{~\bd}(\fb_4)^{\bd}_{~\ad}
\end{eqnarray}
where $f_i\equiv f(p_i,a_i)$. Then using (\ref{eq:f->F}), (\ref{eq:B4}) and (\ref{eq:B4b}) we have
 \begin{eqnarray}
(f_a)_{\a}^{~\b} (f_b)_{\b}^{~\a} & = & -\frac{1}{4}\left( (F^2)_{ab} + i (F_a)^{\m\n}(\wt{F}_b)_{\n\m}\right) \label{eq:2f} \\
(\fb_a)^{\ad}_{~\bd} (\fb_b)^{\bd}_{~\ad} & = & -\frac{1}{4}\left( (F^2)_{ab} - i (F_a)^{\m\n}(\wt{F}_b)_{\n\m}\right) \label{eq:2fb}
\end{eqnarray}
In general we use the notation
\begin{eqnarray}
  \label{eq:F^n}
  (F^n)_{a_1 \ldots a_n} & \equiv & (F_{a_1})^{\m_1}_{~~\m_2} (F_{a_2})^{\m_2}_{~~\m_3} \cdots  (F_{a_n})^{\m_n}_{~~\m_1} 
\label{4complete)}
\end{eqnarray}
with the only exception of
\begin{eqnarray}
(F^2)_{ab} & \equiv & (F_{a})^{\m_1\m_2}(F_{b})_{\m_1\m_2}
\end{eqnarray}
which we keep with indices contracted in standard manner. Making use of (\ref{eq:2tilde1}) and (\ref{eq:2tilde2}) to eliminate $\wt{F}$
the final result becomes
\begin{eqnarray} \label{eq:E4}
&& \int {\rm d}^4 k \frac{1}{k^2 (k+p_2)^2(k+p_2+p_3)^2(k+p_2+p_3+p_4)^2}~g(a_1,a_2,a_3,a_4)\nonumber\\ 
&& \nonumber \\
&& ~~~~~~~\frac{1}{8}\left\{ 4(F^4)_{a_1 a_2 a_3 a_4}+4(F^4)_{a_1 a_2 a_4 a_3}+4(F^4)_{a_1 a_3 a_2 a_4}\right. \nonumber \\
&&~~~~~~~ \left.- (F^2)_{a_1 a_2}(F^2)_{a_3 a_4}- (F^2)_{a_1 a_3}(F^2)_{a_2 a_4}- (F^2)_{a_1 a_4}(F^2)_{a_2 a_3} \right\}
\end{eqnarray}
Now following the general procedure outlined in the previous section  we rewrite (\ref{eq:E4}) as an infinite series of local terms in a low-energy expansion with higher derivatives.  Feynmann combining the propagators in (\ref{box}) we obtain
\begin{eqnarray}
I_4 &=& \int {\rm d}^4 k \frac{1}{k^2 (k+p_2)^2(k+p_2+p_3)^2(k+p_2+p_3+p_4)^2} \nonumber \\
&& \nonumber \\
& \Rightarrow & 3! \int {\rm d}^4 k \int_0^1 {\rm d}x_1 \int_0^{x_1} {\rm d}x_2 \int_0^{x_2} {\rm d}x_3  \nonumber \\
&& \left[k^2 + M^2 +(1-x_1)(p_2+p_3+p_4)^2+(x_1-x_2)(p_2+p_3)^2+(x_2-x_3)(p_2)^2 \right. \nonumber \\
&& \left.-\left((1-x_3)p_2+(1-x_2)p_3+(1-x_1)p_4\right)^2\right]^{-4} \nonumber \\
&& \nonumber \\
&=& \int {\rm d}^4 k \frac{1}{(k^2 + M^2)^4} -4! \int {\rm d}^4 k \int {\rm d}x_i \frac{H_4(x_i,p_j)}{(k^2+M^2)^5} + {\cal O}(\frac{p^4}{M^{8}})
\label{4lenergyexpansion}
\end{eqnarray}
where
\begin{eqnarray}
H_4(x_i,p_j) &=& (1-x_1)(p_2+p_3+p_4)^2+(x_1-x_2)(p_2+p_3)^2+(x_2-x_3)(p_2)^2 \nonumber \\
&&~~~~~~~~~~~~~ -[(1-x_3) p_2+(1-x_2) p_3+(1-x_1) p_4]^2
\label{4Feynmannpar}
\end{eqnarray}
First we look at the leading term in (\ref{4lenergyexpansion}). It does not depend on the external momenta and therefore in (\ref{eq:E4}) we can freely symmetrize on them.  We obtain the $F^4$ non abelian contribution to the Born-Infeld action
\begin{eqnarray}
 \G_4&\Rightarrow &  g(a_1,a_2,a_3,a_4)
\left\{ (F^4)_{a_1 a_2 a_3 a_4}+2(F^4)_{a_1 a_2 a_4 a_3}\right.\nonumber \\
&&~~~~~~~~~~~~~~~~~~ ~~~~~~  ~~~~~~ \left.- \frac{1}{4}[(F^2)_{a_1 a_3}(F^2)_{a_2 a_4}+2 (F^2)_{a_1 a_2}(F^2)_{a_3 a_4}] \right\}\nonumber\\
&=&\Tra\left[F^{\m\n}F_{\n\rho}F^{\rho\s}F_{\s\m}+2F^{\m\n}F_{\n\rho}F_{\s\m}F^{\rho\s}
\right.\nonumber\\
&&\left.~~~~~~~~~~~~-\frac{1}{4}\left(F^{\m\n}F^{\rho\s}F_{\m\n}F_{\rho\s}
+2F^{\m\n}F_{\m\n}F^{\rho\s}F_{\rho\s}\right) \right]
\label{F^4terms}
\end{eqnarray}
We notice that if the gauge group is $SU(N)$ we can express the gauge group matrices which appear in (\ref{F^4terms}) using the standard fundamental representation. If we do so we find that the planar sector (large $N$ limit) reproduces the symmetrized trace structure \cite{Tseytlin1}
\begin{eqnarray}
 \G_4&\Rightarrow &{\rm Tr}\Big( T^{a_1}T^{a_2}T^{a_3}T^{a_4}+ {\rm permutations}\Big)\left\{(F^4)_{a_1a_2a_3a_4} -\frac{1}{4}(F^2)_{a_1a_2}(F^2)_{a_3a_4} \right\} \nonumber \\
& \equiv & STr\left(F^4 -\frac{1}{4}(F^2)^2 \right)
\label{4symmetric}
\end{eqnarray}
Next we study the subleading contribution which corresponds to terms with two derivatives. Including the overall factor $(-1/2)^4 (1/2!)^2 2^3$ from the vertices and combinatorics, the integral containing $H_4(x_i,p_j)$ in
(\ref{4Feynmannpar}) gives
\begin{eqnarray} \label{eq:p^2}
\G_4^{(der)}&=& -\frac{1}{40}\left(\frac{\pi^2}{12}\frac{1}{M^6}\right)\left[10 (p_2)^2 + 8 p_2 \cdot p_3 + 2 p_2 \cdot p_4  \right]~g(a_1,a_2,a_3,a_4) \nonumber \\
&& \frac{1}{8}\left\{\left[ 4(F^4)_{a_1 a_2 a_3 a_4}+4(F^4)_{a_1 a_2 a_4 a_3}+4(F^4)_{a_1 a_3 a_2 a_4}- (F^2)_{a_1 a_2}(F^2)_{a_3 a_4}\right. \right.\nonumber \\
&& \left.\left.- (F^2)_{a_1 a_3}(F^2)_{a_2 a_4}- (F^2)_{a_1 a_4}(F^2)_{a_2 a_3}\right] + {\rm~permutations~}a_2,a_3,a_4 \right\}
\end{eqnarray}
These terms give rise to contributions that are not gauge invariant since from $p\rightarrow i\pa$ ordinary derivatives are produced. The correct covariantization of the result is obtained adding terms with one and two background connections $\mathbf{\Gamma}_{\a\ad}$ from the five- and six-point functions respectively. In the next sections we will compute these terms explicitly. 
\section{The five-point function}
The one-loop contributions to the effective action corresponding to diagrams with five external vector fields group themselves into two distinct classes:
\begin{enumerate}
\item graphs with two ${\bf W}$ and three ${\bf {\Wb}}$ vertices (and corresponding hermitian conjugate ones) which produce gauge-invariant structures;
\item graphs with two ${\bf W}$, two ${\bf{\Wb}}$ vertices and one ${\bf \G}^{\a\ad}\pa_{\a\ad}$ vertex which, as we will show,  contribute to the covariantization of the four-point function.
\end{enumerate}
We start computing the first class of terms.\\

\subsection{The ${\bf W}^2 {\bf {\Wb}}^3$ five-point terms}
Now we consider diagrams with two ${\bf W}$ and three ${\bf {\Wb}}$ vertices as
in Fig.2a. After completion of the $D$-algebra in the loop we obtain terms with five scalar propagators and a typical background dependence of the form 
\begin{equation}
\int d^4\theta~{\bf W}^\a(p_1,a_1){\bf W}_\a(p_2,a_2){\bf \Wb}^\ad(p_3,a_3)\Db_\ad{\bf \Wb}^\bd(p_4,a_4)
{\bf \Wb}_\bd(p_5,a_5)
\end{equation}
\begin{minipage}[r]{0.40 \linewidth}
\includegraphics[width=1 \linewidth]{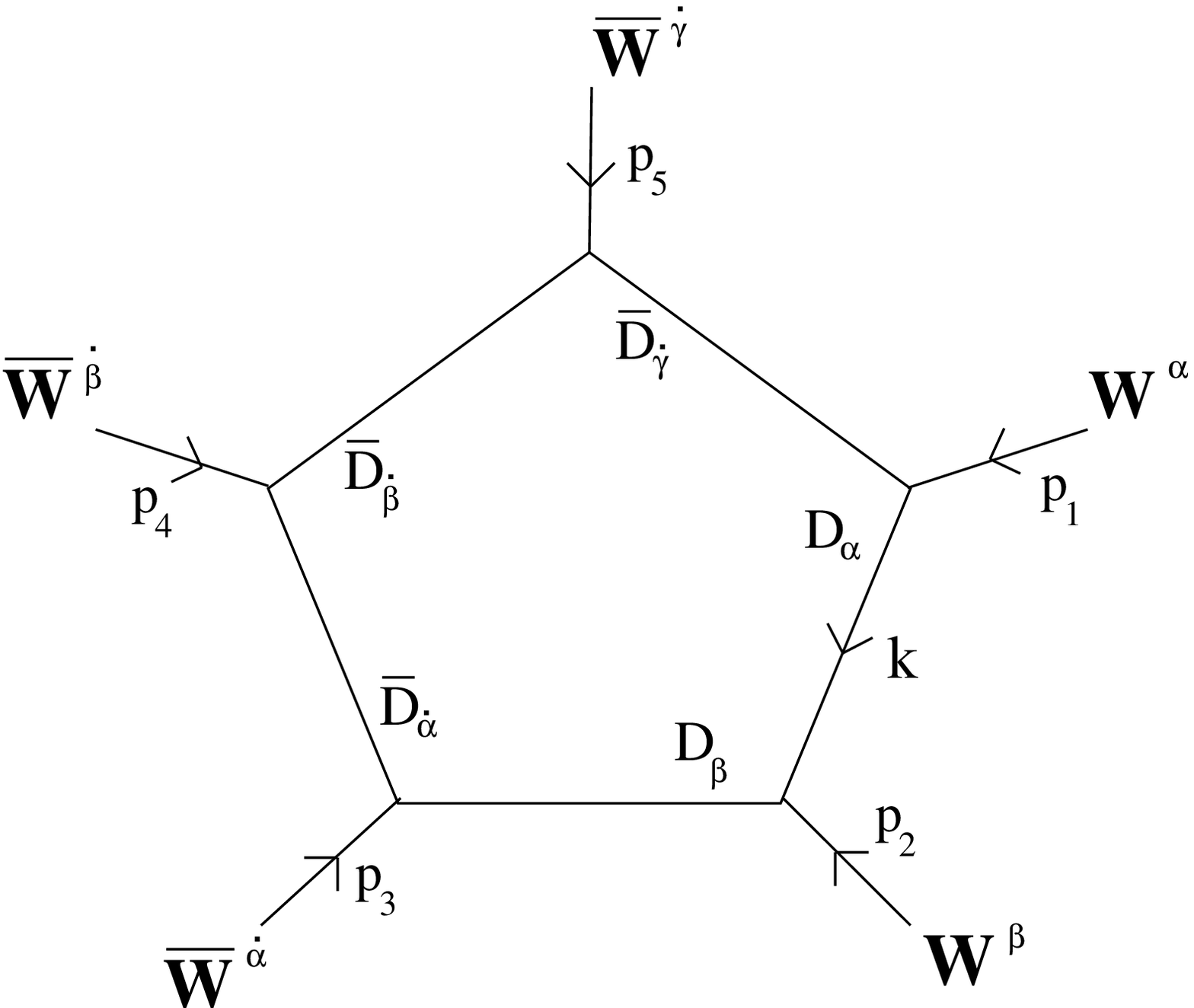}
\begin{center} a) \end{center}
\end{minipage}
~~~~~~~
\begin{minipage}[l]{0.40 \linewidth}
\includegraphics[width=1 \linewidth]{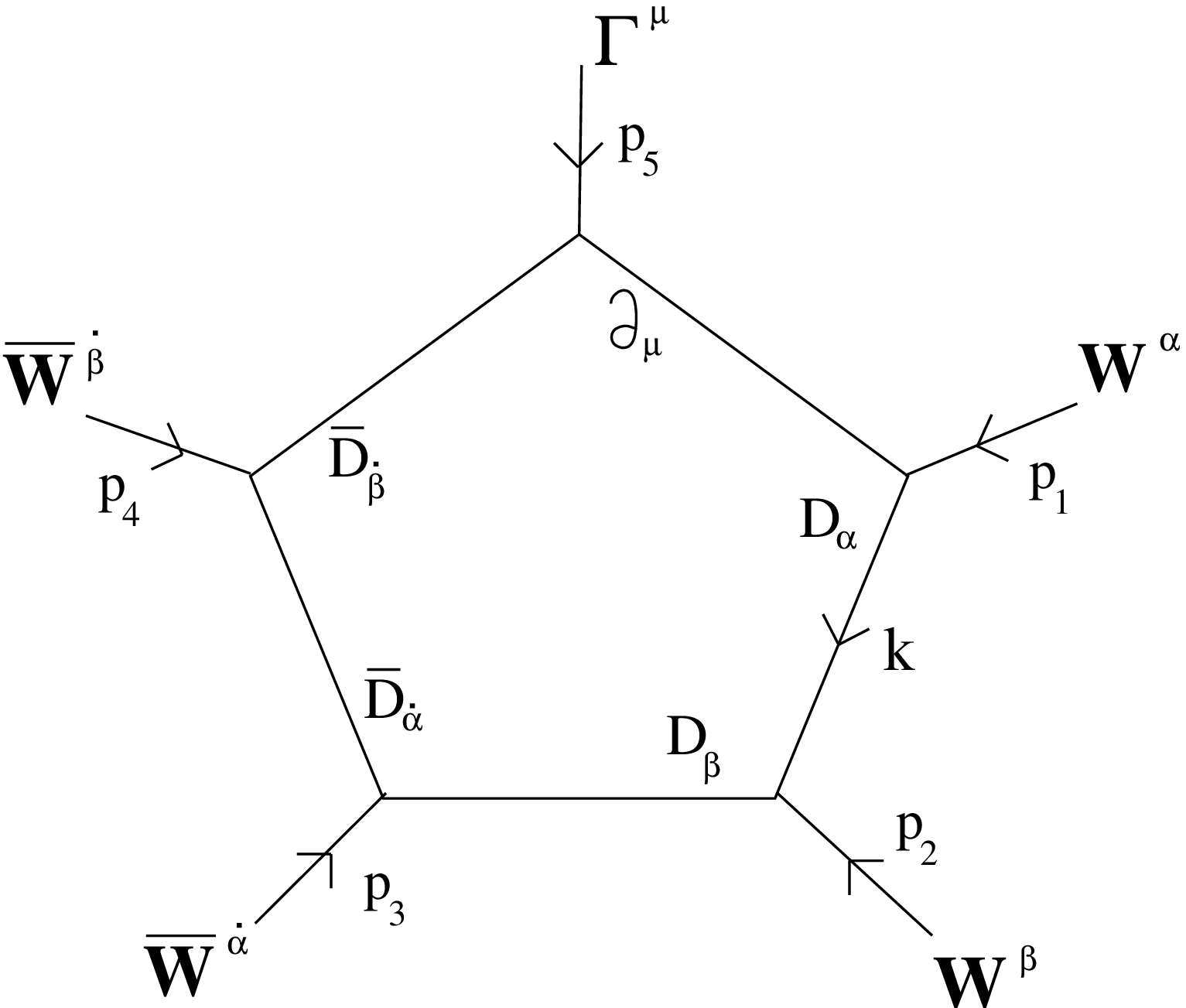}
\begin{center} b) \end{center}
\end{minipage}
\begin{center}
{\small {Figure 2: Five-point amplitudes}}
\end{center}~\\
Including the combinatorial factors and the loop--momentum integration we obtain the corresponding bosonic contributions in terms of the field strengths
\begin{eqnarray}
\G_5^{(1)} &=&  \left(-\frac{1}{2}\right)^5 \frac{1}{2!}\frac{1}{3!} 2^4  ~(f_1)_{\a}^{~\b}(f_2)_{\b}^{~\a}(\fb_3)^{\ad}_{~\bd}
(\fb_4)^{\bd}_{~\gd}(\fb_3)^{\gd}_{~\ad}~g(a_1, a_2, a_3, a_4, a_5) \nonumber \\
\nonumber \\
&& \int {\rm d}^4 k \frac{1}{k^2 (k+p_2)^2(k+p_2+p_3)^2(k+p_2+p_3+p_4)^2(k+p_2+p_3+p_4+p_5)^2} \nonumber\\
&&~~~~~~
\label{G_5^1}
\end{eqnarray}
Once again in order to simplify the notation we define
\begin{eqnarray}
&&(f^n)_{a_1 a_2 \cdots a_n} \equiv (f_{a_1})_{\a}^{~\b}(f_{a_2})_{\b}^{~\g}\cdots(f_{a_n})_{\d}^{~\a} \nonumber\\
&&(\fb^n)_{a_1 a_2 \cdots a_n} \equiv (\fb_{a_1})^{\ad}_{~\bd}(\fb_{a_2})^{\bd}_{~\gd}\cdots
(\fb_{a_n})^{\dd}_{~\ad}
\label{ffbar}
\end{eqnarray}
It is easy to show that $(f^n)_{a_1 a_2 \cdots a_n}$ is symmetric under cyclic permutation of the indices $a_1 a_2 \cdots a_n$ for $n$ even, while it is antisymmetric for $n$ odd.
Using the definitions in (\ref{ffbar}) we denote the contributions corresponding to $\G_5^{(1)}$ simply as $(f^2)_{a_1 a_2}(\fb^3)_{a_3 a_4 a_5}$.
In the same way we obtain another graph interchanging the external background ${\bf W}(p_2,a_2)$ with ${\bf \Wb}(p_3,a_3)$, e.g.
\begin{eqnarray}
\label{G_5^2}
\G_5^{(2)} &=&  \left(-\frac{1}{2}\right)^5 \frac{1}{2!}\frac{1}{3!} 2^4  ~(f^2)_{a_1 a_2}(\fb^3)_{a_3 a_4 a_5}~g(a_1, a_3, a_2, a_4, a_5) \nonumber \\ \nonumber \\
&&\int {\rm d}^4 k \frac{1}{k^2 (k+p_3)^2(k+p_2+p_3)^2(k+p_2+p_3+p_4)^2(k+p_2+p_3+p_4+p_5)^2} \nonumber\\
&&~~~~
\end{eqnarray}
In addition to (\ref{G_5^1}) and (\ref{G_5^2}) we have graphs obtained through a permutation of couples of indices $(p_i,a_i),~ i=3,4,5$ and $(p_1,a_1)(p_2,a_2)$. In this way we end up with a total of $3! \cdot 2! \cdot 2=4!$ inequivalent contributions.
The final result can be rewritten in terms of the field strengths $F_{\m\n}$. Indeed using the formulas (\ref{eq:B6}) e (\ref{eq:Bb6}) in Appendix \ref{s-algebra} we can express the product of three $\fb$'s (or equivalently three $f$'s) as
\begin{eqnarray}
  \label{eq:fb^3bis}
(\fb_a)^{\ad}_{~\bd} (\fb_b)^{\bd}_{~\gd} (\fb_c)^{\gd}_{~\ad} &=& - \frac{1}{8}(2 (F^3)_{abc} + i (F_a)^{\m}_{~\n}(F_b)^{\n}_{~\s}(\wt{F}_c)^{\s}_{~\m}+ i (\wt{F}_a)^{\m}_{~\n}(F_b)^{\n}_{~\s}(F_c)^{\s}_{~\m}) \nonumber \\
\end{eqnarray}
In the same way using (\ref{eq:B4}) and (\ref{eq:B4b}) two $\fb$'s (or two $f$'s) can be rewritten as in \eqn{eq:2f} and \eqn{eq:2fb}.
The product of \eqn{eq:2f} and \eqn{eq:fb^3bis} gives
\begin{eqnarray}
&& \frac{1}{32}\left[(F^2)_{a_1 a_2} + i (F_{a_1})^{\m\n}(\wt{F}_{a_2})_{\n\m}\right]\nonumber\\
&&~~~~~~~~~~~~
\left[2 (F^3)_{a_3a_4a_5} + i (F_{a_3})^{\m}_{~\n}(F_{a_4})^{\n}_{~\s}(\wt{F}_{a_5})^{\s}_{~\m}+ i (\wt{F}_{a_3})^{\m}_{~\n}(F_{a_4})^{\n}_{~\s}(F_{a_5})^{\s}_{~\m}\right] 
\label{intermediate}
\end{eqnarray}
Finally using (\ref{eq:2tilde1}) in order to reexpress the two $\wt{F}$ and adding the hermitian conjugate contributions, we obtain
\begin{eqnarray}
\label{eq:risulta5}
\Re\left((f^2)_{a_1 a_2}(\fb^3)_{a_3 a_4 a_5}\right) &=& \frac{1}{32} \left[2 F^5_{a_1 a_4 a_3 a_2 a_5} - 2 F^5_{a_1 a_4 a_5 a_2 a_3} + 2 F^5_{a_1 a_5 a_4 a_2 a_3} \right.  \nonumber\\
&& -  2 F^5_{a_1 a_3 a_4 a_2 a_5} + 2 F^2_{a_1 a_2}F^3_{a_3 a_4 a_5} - F^2_{a_1 a_5}F^3_{a_2 a_3 a_4}  \nonumber\\
&&\left.-F^2_{a_2 a_5}F^3_{a_1 a_3 a_4} -F^2_{a_1 a_3}F^3_{a_2 a_4 a_5} -F^2_{a_2 a_3}F^3_{a_1 a_4 a_5}\right] 
\end{eqnarray} 
Now, as we have done in the previous section, we study the loop momentum integral in the low-energy approximation
\begin{eqnarray}
I_5 & = &\int {\rm d}^4 k \frac{1}{k^2 (k+p_2)^2(k+p_2+p_3)^2(k+p_2+p_3+p_4)^2(k+p_2+p_3+p_4+p_5)^2} \nonumber \\
&& \nonumber \\
&\Rightarrow& 4! \int {\rm d}^4 k \int_0^1 {\rm d}x_1 \int_0^{x_1} {\rm d}x_2 \int_0^{x_2} {\rm d}x_3 \int_0^{x_3} {\rm d}x_4 \nonumber \\
&& \left[k^2 + M^2 +(1-x_1)(p_2+p_3+p_4+p_5)^2+(x_1-x_2)(p_2+p_3+p_4)^2 \right. 
\nonumber \\
&& +(x_2-x_3)(p_2+p_3)^2+ (x_3-x_4)(p_2)^2  -(1-x_4)^2(p_2)^2-(1-x_3)^2(p_3)^2 \nonumber \\
&&\left.~~~~~~~~~~ -(1-x_2)^2(p_4)^2-(1-x_1)^2(p_5)^2\right]^{-5}\nonumber\\
&& \nonumber \\
&=& \int {\rm d}^4 k \frac{1}{(k^2 + M^2)^5} -5! \int {\rm d}^4 k \int {\rm d}x_i \frac{H_5(x_i,p_j)}{(k^2+M^2)^6} + {\cal O}(\frac{p^4}{M^{10}})
\label{lenergy5point} 
\end{eqnarray}
where
\begin{eqnarray}
H_5(x_i,p_j) &=& (1-x_1)(p_2+p_3+p_4+p_5)^2+(x_1-x_2)(p_2+p_3+p_4)^2 \nonumber \\
&& +(x_2-x_3)(p_2+p_3)^2 + (x_3-x_4)(p_2)^2 \nonumber \\
&& -[(1-x_4)(p_2)+(1-x_3)(p_3)+(1-x_2)(p_4)+(1-x_1)(p_5)]^2 
\label{Feynmannpar5point}
\end{eqnarray}
We are interested only in the leading order contributions which again are independent of the external momenta $p_1,\ldots,p_5$, so that the various terms can be combined as
\begin{eqnarray}
\G_{5}^{(tot)} &=& -\frac{1}{12}\left(\int{\rm d}^4 k \frac{1}{(k^2 + M^2)^5}\right) ~\Re\left((f^2)_{a_1 a_2}(\fb^3)_{a_3 a_4 a_5}\right) \left[ g(a_1 \left\{a_2,a_3\right\}, a_4, a_5)\right. \nonumber \\
&& \nonumber \\
&&~~~~~~~~~~~\left.+ {~~ \rm permutations~~of~~ } a_1,a_2 ~~ {\rm and} ~~ a_3,a_4,a_5\right] 
\end{eqnarray}
Taking into account the symmetry properties of $(f^n)_{a_1 a_2 \cdots a_n}$, we obtain
\begin{eqnarray}
\G_{5}^{(tot)}&=& -\frac{1}{12}\left(\int {\rm d}^4 k \frac{1}{(k^2 + M^2)^5}\right) ~\Re\left((f^2)_{a_1 a_2}(\fb^3)_{a_3 a_4 a_5}\right)  \\ \nonumber \\
&& 2\left[ g(a_1 \left\{a_2,a_3\right\} \left[a_4, a_5\right]) +g(a_1 \left\{a_2,a_4\right\} \left[a_5, a_3\right])+g(a_1 \left\{a_2,a_5\right\} \left[a_3, a_4\right])\right] \nonumber\\
&&~~~~\nonumber
\end{eqnarray}
Finally we can use the result in \eqn{eq:risulta5} and perform explicitly the commutator algebra in the gauge group structures
\begin{eqnarray}
\label{eq:5puntifinale}
\G_{5}^{(tot)}&=& -\frac{1}{32}\left(\frac{\pi^2}{12}\frac{1}{M^6}\right) ~ g(a_1,a_2,a_3,a_4,a_5) \nonumber \\ \nonumber \\
&& [-2 (F^5)_{a_1a_2a_3a_4a_5}
-(F^5)_{a_1a_4a_2a_5a_3}
+3(F^5)_{a_1a_4a_3a_2a_5} \nonumber \\ \nonumber \\
&& +1/2 (F^2)_{a_2a_4}(F^3)_{a_1a_3a_5}-1/2 (F^2)_{a_3a_4}(F^3)_{a_1a_2a_5}]\nonumber\\
&&~~~~~
\end{eqnarray}
The above expression can be manipulated further so that the  $F^2 F^3$ terms are eliminated in favor of the $F^5$ terms, using the identities (\ref{eq:identities}) depicted graphically in Fig.5.
\begin{eqnarray}
\G_{5}^{(tot)}&=& -\frac{1}{32}\left(\frac{\pi^2}{12}\frac{1}{M^6}\right) ~ g(a_1,a_2,a_3,a_4,a_5) \nonumber \\ \nonumber \\
&& \left[-\frac{8}{5} (F^5)_{a_1a_2a_3a_4a_5}
-\frac{4}{5}(F^5)_{a_1a_4a_2a_5a_3}
+4(F^5)_{a_1a_4a_3a_2a_5}\right]
\label{final5point}
\end{eqnarray}
In section 6 we will come back to (\ref{final5point}) and assemble all our results.
Now we turn to the computation of the second class of diagrams that contribute to the five-point function.
\subsection{The ${\bf W}{\bf W}{\bf \Wb}{\bf \Wb} {\bf \G}^{\g\gd} \pa_{\g\gd}$ five-point terms} 
In this subsection we analyze contributions from the five-point function with a background dependence of the form ${\bf W}{\bf W}{\bf \Wb}{\bf \Wb} {\bf \G}^{\g\gd}$. These terms are not gauge invariant, but as we have previously stated, they do contribute to the covariantization of derivative terms obtained from the four-point function. In order to prove this we start considering a typical graph of this type as shown in Fig.2b.
After completion of the $D$-algebra  it gives rise to
\begin{eqnarray}
&&\int d^4\theta~{\bf \G}_{\g\gd}(p_5,a_5){\bf W}^\a(p_1,a_1){\bf W}_\a(p_2,a_2){\bf \Wb}^\ad(p_3,a_3)
{\bf \Wb}_\ad(p_4,a_4)\nonumber\\
&&~~~~~ \Rightarrow ({\bf \G}_5)_{\g\gd}(f_1)_{\a}^{~\b}(f_2)_{\b}^{~\a}(\fb_3)^{\ad}_{~\bd}(\fb_4)^{\bd}_{~\ad} 
\end{eqnarray}
Including the loop-momentum integration we find
\begin{eqnarray}
\G_5^{(cov1)} &=& \frac{1}{8}\int {\rm d}^4 k \frac{-i(k-p_1)_{\g\gd}}{k^2 (k+p_2)^2(k+p_2+p_3)^2(k+p_2+p_3+p_4)^2(k-p_1)^2} \nonumber \\
&& \nonumber \\
&& ~~~~~~~~[g(a_1,a_2,a_3,a_4,a_5) ({\bf \G}_5)^{\g\gd}(f_1)_{\a}^{~\b}(f_2)_{\b}^{~\a}(\fb_3)^{\ad}_{~\bd}(\fb_4)^{\bd}_{~\ad}]
\label{2W2WbarGamma}
\end{eqnarray}
The complete set of diagrams is obtained considering all the permutations of the indices $(p_1,a_1)\cdots(p_4,a_4)$.
The next step is to evaluate the momentum integral to leading order in the low-energy expansion
\begin{eqnarray}
I_5^{(cov1)} & \Rightarrow & 24 \int {\rm d}^4 k \int {\rm d}x_i \frac{1}{(k^2 +M^2)^5} \nonumber\\
&&~~~~~ \left[p_2 (x_4-x_1) + p_3 (x_3-x_1)+ p_4 (x_2-x_1) - x_1 p_1\right]_{\g\gd}
\end{eqnarray}
The final result is
\begin{eqnarray} \label{eq:5punti1gamma}
\G_5^{(cov1)} &=& -\frac{i}{40} \left(\frac{\pi^2}{12}\frac{1}{M^6}\right)  g(a_1,a_2,a_3,a_4,a_5)
\left[4 p_1 + 3 p_2 + 2 p_3 + p_4 \right]_{\g\gd} \\
&& \left[({\bf \G}_5)^{\g\gd}(f_1)_{\a}^{~\b}(f_2)_{\b}^{~\a}(\fb_3)^{\ad}_{~\bd}(\fb_4)^{\bd}_{~\ad} + ~{\rm~ permutations~ of ~}(p_1,a_1)\cdots(p_4,a_4)\right] \nonumber
\end{eqnarray}
Now we want to show that (\ref{eq:5punti1gamma}) gives the correct contribution linear in ${\bf \G}$ to covariantize (\ref{eq:p^2}). (The part quadratic in 
${\bf \G}$ is produced at the level of the six-point function and it will be discussed in the next section.)
Thus we go back to (\ref{eq:p^2}) and there we substitute every ordinary  derivative with a covariant derivative whose expression in momentum space can be written as
\begin{eqnarray}
(\DC \psi)^c(p)&=& (\pa \psi)^c(p) +  \int {\rm d}^4 q~ ({\bf \G}_a)(q)~\psi_b(p-q) f^{abc}
\label{covdermom}
\end{eqnarray} 
In so doing we find that the terms which are linear in ${\bf \G}$ exactly amount to the ones in \eqn{eq:5punti1gamma}. This can be checked in a simple manner as follows: first we write the derivative contributions to the four-point function as
\begin{eqnarray} \label{C_4der-bis}
(\G_4)_{der} & = &-\frac{1}{40}\left(\frac{\pi^2}{12}\frac{1}{M^6}\right) \left[10 (p_2)^2 + 8 p_2 \cdot p_3 + 2 p_2 \cdot p_4   \right] \nonumber \\
&& \frac{1}{4}[\Tra(A(p_1)B(p_2)C(p_3)D(p_4)) + ~{\rm permutations~of~ }A,B,C,D]
\end{eqnarray}
where $A,B,C,D$ denote either $f_{\a\b}$ or $\fb_{\ad\bd}$, with all permutations included while keeping the momenta in the order $p_1,p_2,p_3,p_4$. Then we can use the ciclicity and inversion property of the trace so that we always have the two derivatives acting both on the first field or on the first and the second or on the first and the third. In this way the rules for the covariantization are simply given by (here we consider the terms linear in ${\bf \G}$ only)
\begin{eqnarray}
  \label{eq:covariantizzazione1}
\Tra(\pa^2 ABCD) &\Rightarrow & 2 \Tra([{\bf \G},\pa A]BCD) \\
& \Rightarrow & 2 \Tra({\bf \G} \pa A BCD) - 2 \Tra({\bf \G} BCD \pa A) \nonumber \\
&& \nonumber \\
\Tra(\pa A \pa BCD) & \Rightarrow & \Tra(\pa A [{\bf \G},B]CD)+\Tra([{\bf \G},A] \pa BCD) \nonumber \\
& \Rightarrow &\Tra({\bf \G} BCD \pa A) - \Tra({\bf \G} CD \pa AB) \nonumber \\
&& + \Tra({\bf \G} A\pa BCD) - \Tra({\bf \G} \pa B CDA) \nonumber \\
&& \nonumber \\
\Tra(\pa A  B \pa CD) & \Rightarrow & \Tra(\pa A B[{\bf \G},C]D)+\Tra([{\bf \G},A]  B\pa CD) \nonumber \\
& \Rightarrow &\Tra({\bf \G} CD \pa AB) - \Tra({\bf \G} D \pa ABC) \nonumber \\
&& + \Tra({\bf \G} AB\pa CD) - \Tra({\bf \G} B\pa  CDA) \nonumber
\end{eqnarray}
In order to streamline the notation we have written the above expressions in the gauge $\pa_{\g\gd} {\bf \G}^{\g\gd}=0$, a choice actually not necessary to prove the result. 
Since we have to include a sum on all permutations we can freely rename the fields and write the final result in the form
\begin{eqnarray}
(\G_4)_{cov1} &=& -\frac{i}{40}\left(\frac{\pi^2}{12}\frac{1}{M^6}\right) \left[4 p_1 + 3 p_2 + 2 p_3 +p_4 \right]_{\g\gd}
[\Tra({\bf \G}^{\g\gd}(p_5)A(p_1)B(p_2)C(p_3)D(p_4))\nonumber\\
&&~~~ +~ {\rm permutations~of~ }A,B,C,D]
\label{finalcovlin}
\end{eqnarray}
which exactly reproduces the formula \eqn{eq:5punti1gamma}.
\section{${\bf \G}{\bf \G}$ contributions from the six-point function}
Now we study contributions from the six-point function containing a ${\bf \G}{\bf \G}$ background dependence which will complete the covariantization of the two-derivative part of the four-point function. There are two distinct types of diagrams as shown in Fig.3.
We consider them in the next two subsections.\\

\subsection{Diagrams with a vertex ${\bf \G}^{\g\gd}{\bf \G}_{\g\gd}$}
The typical background dependence is given by
\begin{eqnarray}
&&\int d^4\theta~ {\bf \G}^{\g\gd}(p_5,a_5){\bf \G}_{\g\gd}(p_6,a_6){\bf W}^\a(p_1,a_1){\bf W}_\a(p_2,a_2){\bf \Wb}^\ad(p_3,a_3)
{\bf \Wb}_\ad(p_4,a_4)\nonumber\\
&&~~~~ \Rightarrow ({\bf \G}_5)^{\g\gd} ({\bf \G}_6)_{\g\gd}(f_1)_{\a}^{~\b}(f_2)_{\b}^{~\a}(\fb_3)^{\ad}_{~\bd}(\fb_4)^{\bd}_{~\ad} \end{eqnarray}
\begin{minipage}[l]{0.45 \linewidth}
\includegraphics[width=1 \linewidth]{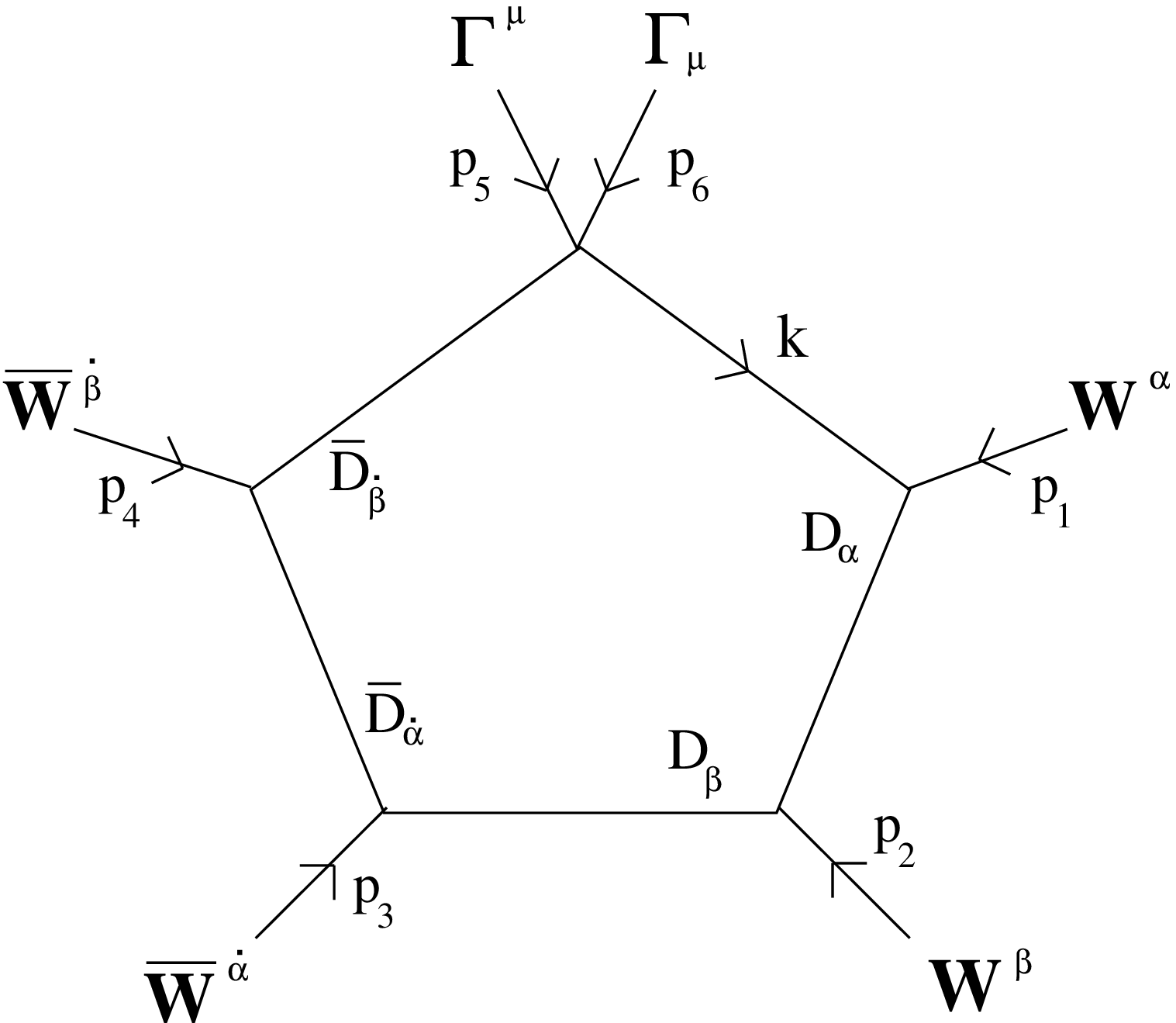}
\begin{center} a) \end{center}
\end{minipage}
~~~~~~~~~~
\begin{minipage}[r]{0.40 \linewidth}
\includegraphics[width=1 \linewidth]{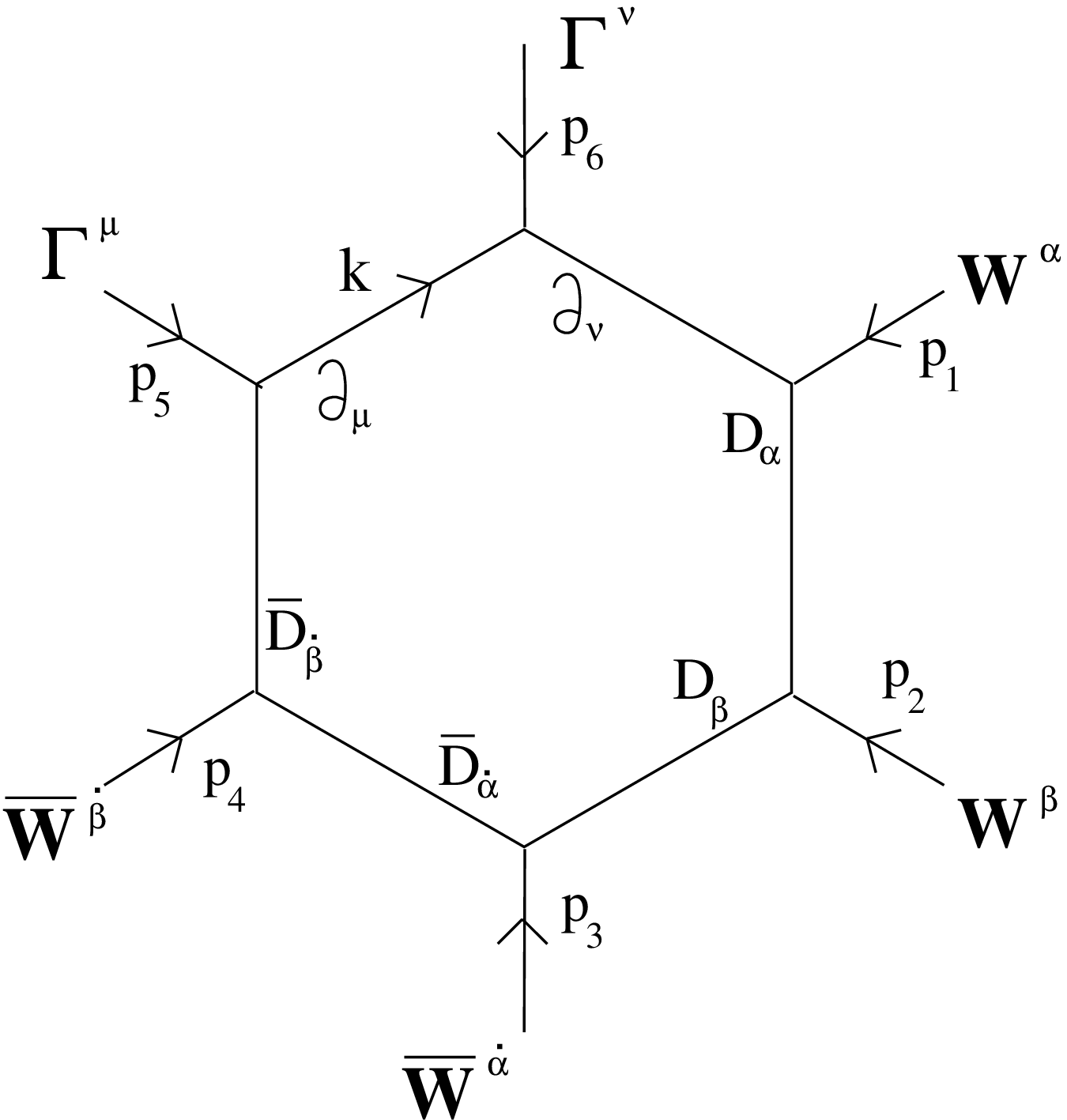}
\begin{center} b) \end{center}
\end{minipage}
\begin{center}
{\small {Figure 3: Six-point amplitude}}
\end{center}
Including the momentum integration one finds
\begin{eqnarray}
&& \G_6^{{\bf \G}{\bf \G}} = -(-\frac{1}{2})^4(-\frac{1}{4})(\frac{1}{2!})^2 2^4 \nonumber\\ &&\left[g(a_1,a_2,a_3,a_4,a_5,a_6) ({\bf \G}_5)^{\g\gd} ({\bf \G}_6)_{\g\gd}(f_1)_{\a}^{~\b}(f_2)_{\b}^{~\a}(\fb_3)^{\ad}_{~\bd}(\fb_4)^{\bd}_{~\ad}\right] \nonumber \\
&& ~~\int {\rm d}^4 k \frac{1}{k^2 (k+p_1)^2(k+p_1+p_2)^2(k+p_1+p_2+p_3)^2(k+p_1+p_2+p_3+p_4)^2} 
\end{eqnarray}
To the above contribution we have to sum all the ones obtained permuting the $a_1,\dots,a_4$ indices. Once again we compute the momentum integral at the leading order in the low-energy expansion. We have
\begin{eqnarray} \label{eq:6punti2gamma}
&&\G_6^{{\bf \G}{\bf \G}} \Rightarrow \frac{1}{16} \left(\frac{\pi^2}{12}\frac{1}{M^6}\right)  g(a_1,a_2,a_3,a_4,a_5,a_6) \nonumber \\
&&~~~ [({\bf \G}_5)^{\g\gd}({\bf \G}_6)_{\g\gd} (f_1)_{\a}^{~\b}(f_2)_{\b}^{~\a}(\fb_3)^{\ad}_{~\bd}(\fb_4)^{\bd}_{~\ad} + {\rm permutations~ of~ }(p_1,a_1)\cdots(p_4,a_4)] \nonumber \\
&&= \frac{1}{16} \left(\frac{\pi^2}{12}\frac{1}{M^6}\right) ({\bf \G}_5)^{\g\gd}({\bf \G}_6)_{\g\gd} (f_1)_{\a}^{~\b}(f_2)_{\b}^{~\a}(\fb_3)^{\ad}_{~\bd}(\fb_4)^{\bd}_{~\ad} \nonumber \\
&& ~~~~[g(a_1,a_2,a_3,a_4,a_5,a_6) + {\rm permutations~ of~ }a_1\cdots a_4]
\end{eqnarray}
Notice that the permutations are only on the indices of the ${\bf W}$'s and the  ${\bf \Wb}$'s.
\subsection{Diagrams $({\bf \G}^{\g\gd}\pa_{\g\gd})({\bf \G}^{\d\dd}\pa_{\d\dd}){\bf W}{\bf W}{\bf \Wb}{\bf \Wb}$}
Depending on the relative position of the two vertices ${\bf \G}^{\g\gd} \pa_{\g\gd}$ one obtains three different diagrams.
The first one contains the two adjacent vertices and produces a contribution of the form
\begin{eqnarray}
\G_{6}^{{\bf \G}\pa {\bf \G}\pa} &=& -
(-\frac{1}{2})^6 (\frac{1}{2!})^3 2^5 \int {\rm d}^4 k ~~ ({\bf \G}_5)^{\g\gd} ({\bf \G}_6)^{\d\dd} (f_1)_{\a}^{~\b}(f_2)_{\b}^{~\a} (\fb_3)^{\ad}_{~\bd}(\fb_4)^{\bd}_{~\ad}  \nonumber \\
&& \frac{k_{\g\gd} (k+p_6)_{\d\dd}}{k^2 (k+p_6)^2(k+p_6+p_1)^2(k+p_6+p_1+p_2)^2(k-p_5-p_4)^2(k-p_5)^2} \nonumber \\
&&  g(a_1,a_2,a_3,a_4,a_5,a_6)
\end{eqnarray}
The leading order in the low-energy expansion gives
\begin{eqnarray}
\G_{6}^{{\bf \G}\pa {\bf \G}\pa} & \Rightarrow & -\frac{5!}{16} \int {\rm d}^4 k ~\int_0^1 {\rm d}x_1 \cdots \int_0^{x_4} {\rm d}x_{5}~ ({\bf \G}_5)^{\g\gd} ({\bf \G}_6)^{\d\dd} (f_1)_{\a}^{~\b}(f_2)_{\b}^{~\a} (\fb_3)^{\ad}_{~\bd}(\fb_4)^{\bd}_{~\ad} \nonumber \\
&&\frac{1}{(k^2+M^2)^6}~ g(a_1,a_2,a_3,a_4,a_5,a_6)\\
&&  \left[k_{\g\gd} k_{\d\dd} + 
(p_6(1-x_5)+p_1(1-x_4)+p_2(1-x_3)+p_3(1-x_2)+p_4(1-x_1))_{\g\gd}\right. \nonumber \\
&& \left.(p_6(1-x_5)+p_1(1-x_4)+p_2(1-x_3)+p_3(1-x_2)+p_4(1-x_1)-p_6)_{\d\dd}\right] 
\nonumber
\end{eqnarray}
From the above contribution we need extract the terms proportional to
 $k_{\g\gd} k_{\d\dd} \sim \frac{1}{2} k^2 C_{\g\d}C_{\gd\dd}$ which are the ones relevant for the covariatization of the four-point function.

The other two diagrams lead to similar contributions. We end up with the following result
\begin{eqnarray}\label{eq:6punti2gammadelta}
\G_{6~tot}^{{\bf \G}\pa {\bf \G}\pa} &=& -\frac{1}{32} \left(\frac{\pi^2}{30}\frac{1}{M^6}\right) ({\bf \G}_5)^{\g\gd} ({\bf \G}_6)_{\g\gd} (f_1)_{\a}^{~\b}(f_2)_{\b}^{~\a} (\fb_3)^{\ad}_{~\bd}(\fb_4)^{\bd}_{~\ad} \nonumber \\
&&  \left[\Big(2g(a_1,a_2,a_3,a_4,a_5,a_6)+2g(a_1,a_6,a_2,a_3,a_4,a_5) \right. \nonumber \\
&&\left. +g(a_1,a_2,a_6,a_3,a_4,a_5)\Big) + {\rm permutations~of~}a_1,a_2,a_3,a_4\right]
\end{eqnarray}
Notice that the first two graphs have a multiplicity which is double as compared to the third one.  All together they reconstruct the $5!$ permutations of the $a_2,\ldots,a_6$ indices.
\subsection{Covariantization of $\G_{4}$ with two ${\bf \G}$'s}
The starting point is the result in \eqn{C_4der-bis}. We proceed in complete analogy with what we have done for the part linear in ${\bf \G}$, so that we obtain simple rules also for the terms quadratic  in ${\bf \G}$:
\begin{eqnarray}
  \label{eq:covariantizzazione2}
\Tra(\pa^2 ABCD) &\Rightarrow &  \Tra([{\bf \G},[{\bf \G}, A]]BCD) \nonumber\\
& \Rightarrow &  \Tra({\bf \G}{\bf \G} A BCD) -  2\Tra({\bf \G} A {\bf \G} BCD)+\Tra({\bf \G}{\bf \G} BCDA) \nonumber \\
&& \nonumber \\
\Tra(\pa A \pa BCD) & \Rightarrow & \Tra([{\bf \G},A] [{\bf \G},B]CD) \nonumber \\
& \Rightarrow &\Tra({\bf \G} A{\bf \G} BCD) - \Tra({\bf \G} AB{\bf \G} CD) \nonumber \\
&& - \Tra({\bf \G}{\bf \G} BCDA) + \Tra({\bf \G} B {\bf \G} CDA) \nonumber \\
&& \nonumber \\
\Tra(\pa A  B \pa CD) & \Rightarrow & \Tra([{\bf \G},A] B[{\bf \G},C]D) \nonumber \\
& \Rightarrow &\Tra({\bf \G} AB {\bf \G} CD)-\Tra({\bf \G} AB C {\bf \G} D) \nonumber \\
&& -\Tra({\bf \G} B {\bf \G} CDA)+\Tra({\bf \G} B C {\bf \G} DA)
\end{eqnarray}
Once again, using the properties of the trace operation it is possible to set one of the ${\bf \G}$'s in first position, and the other one either in second or in third or in fourth position. Then renaming the fields and  using the formulas (\ref{kintegrals}) we obtain
\begin{eqnarray}
(\G_4)_{cov2} &=& -\frac{1}{80}\left(\frac{\pi^2}{12}\frac{1}{M^6}\right) [-3\Tra({\bf \G}(p_5){\bf \G}(p_6)A(p_1)B(p_2)C(p_3)D(p_4))\\ 
&& ~~~~~+2 \Tra({\bf \G}(p_5)A(p_1){\bf \G}(p_6)B(p_2)C(p_3)D(p_4) \nonumber \\
&& ~~~~~+{\bf \G}(p_5)A(p_1)B(p_2){\bf \G}(p_6)C(p_3)D(p_4))  + \textrm{ permutations~of~ }A,B,C,D]\nonumber
\end{eqnarray}
which is just the sum of \eqn{eq:6punti2gamma} and of \eqn{eq:6punti2gammadelta}.
This completes the proof of the covariantization of the two-derivative terms for the four-point function.
\section{The final result}
In this section we collect all the contributions at order $M^{-6}$ that we have computed so far, i.e. the two derivative terms from the first subleading contribution of the four-point function in (\ref{eq:p^2}), and the leading terms of the five-point function in (\ref{final5point}). We start reexamining the two derivative terms.

Covariantizing the expression in (\ref{eq:p^2}) and taking into account the cyclic and inversion properties of the trace operation (cfr.(\ref{colourstructure},\ref{trace})), we have
\begin{eqnarray}
(\Gamma_4)_{der} & = & -\frac{1}{8}\left(\frac{\pi^2}{12}\frac{1}{M^6}\right) \Tra \left\{
-\DC^2 F_{\m\n}F_{\n\rho}F_{\rho\s}F_{\s\m}
-2 \DC^2 F_{\m\n}F_{\rho\s}F_{\n\rho}F_{\s\m}
 \right. \nonumber \\
&& +\frac{1}{2}\DC^2 F_{\m\n}F_{\m\n}F_{\rho\s}F_{\rho\s}
+\frac{1}{4}\DC^2 F_{\m\n}F_{\rho\s}F_{\m\n}F_{\rho\s}
-\frac{4}{5}\DC F_{\m\n} \DC F_{\n\rho}F_{\rho\s}F_{\s\m}
 \nonumber \\
&&-\frac{4}{5}\DC F_{\m\n} \DC F_{\rho\s}F_{\n\rho}F_{\s\m}
-\frac{1}{5}\DC F_{\m\n}F_{\n\rho}\DC F_{\rho\s}F_{\s\m}
-\frac{2}{5}\DC F_{\m\n}F_{\rho\s}\DC F_{\n\rho}F_{\s\m} 
\nonumber \\
&& -\frac{4}{5} F_{\m\n}\DC F_{\rho\s} \DC F_{\n\rho}F_{\s\m} 
+\frac{1}{5}\DC F_{\m\n} \DC F_{\m\n}F_{\rho\s}F_{\rho\s}
+\frac{1}{5} F_{\m\n}\DC F_{\m\n}\DC F_{\rho\s}F_{\rho\s}
 \nonumber \\
&& \left.+\frac{1}{5}\DC F_{\m\n}\DC F_{\rho\s}F_{\m\n}F_{\rho\s}
+\frac{1}{10} F_{\m\n}\DC F_{\m\n}F_{\rho\s}\DC F_{\rho\s}
+\frac{1}{20}F_{\m\n}\DC F_{\rho\s}F_{\m\n}\DC F_{\rho\s} \right\} \nonumber \\
\end{eqnarray}
where it is understood that the indices on the two $\nabla$ are to be contracted.
We integrate by parts in such a way to have the derivatives acting either on the same or on two adjacent field strengths. We obtain \\
\begin{minipage}[l]{1 \linewidth}
\begin{eqnarray} \label{eq:p^2finale}
&& \!\!\!\!\!\!\! (\G_4)_{der}= \nonumber \\
&& \frac{1}{20}\left(\frac{\pi^2}{12}\frac{1}{M^6}\right) \Tra \Big\{
(\DC F_{\m\n} \DC F_{\n\rho}F_{\rho\s}F_{\s\m}
+\DC F_{\m\n} \DC F_{\rho\s}F_{\n\rho}F_{\s\m}+ F_{\m\n}\DC F_{\rho\s} \DC F_{\n\rho}F_{\s\m})
 \nonumber \\
&&\left. -\frac{1}{4}(\DC F_{\m\n} \DC F_{\m\n}F_{\rho\s}F_{\rho\s}
+F_{\m\n}\DC F_{\m\n}\DC F_{\rho\s}F_{\rho\s}
+\DC F_{\m\n}\DC F_{\rho\s}F_{\m\n}F_{\rho\s})\right\} \nonumber \\
&& \nonumber \\
&& -\frac{1}{20}\left(\frac{\pi^2}{12}\frac{1}{M^6}\right) \Tra \left\{
-2\DC^2 F_{\m\n}F_{\n\rho}F_{\rho\s}F_{\s\m}
-4 \DC^2 F_{\m\n}F_{\rho\s}F_{\n\rho}F_{\s\m} \right. \nonumber \\
&& \left.~~~~~~~~~+ \DC^2 F_{\m\n}F_{\m\n}F_{\rho\s}F_{\rho\s} 
+\frac{1}{2}\DC^2 F_{\m\n}F_{\rho\s}F_{\m\n}F_{\rho\s} \right\}
\end{eqnarray}
\end{minipage}
\vskip 3mm
Now we show that quite generally we can rewrite terms which contain  $\DC^2$ and four $F$'s  as $F^5$ contributions. Indeed, using the equations of motion $\nabla^\m F_{\m\n}=0$ and the Bianchi identities $ \nabla_{[\m} F_{\n\rho]}=0$,  we have the following identities
\begin{eqnarray}\label{eq:D^2}
\DC^2 F_{\m\n}&=& \DC_{\s}\DC_{\s} F_{\m\n} \stackrel{\textrm{\tiny Bianchi}}{=} -\DC_{\s}(\DC_{\m} F_{\n\s}+\DC_{\n} F_{\s\m})\nonumber\\
&&~~\nonumber\\
& \stackrel{e.o.m}{=} & -[\DC_{\s},\DC_{\m}]F_{\n\s}-[\DC_{\s},\DC_{\n}] F_{\s\m}\nonumber\\
&&~~\nonumber\\
&=& +\frac{i}{2} [F_{\s\m},F_{\n\s}]+\frac{i}{2}[F_{\s\n},F_{\s\m}]= i (F_{\m\s}F_{\s\n}-F_{\n\s}F_{\s\m})
\end{eqnarray}
where we have used $[\DC_{\m},\DC_{\n}]F_{\rho\s}=-\frac{i}{2}[F_{\m\n},F_{\rho\s}]$.\\
Therefore we can write
\begin{equation}
\DC^2 (F^a)_{\m\n}=- f_{abc}(F^b)_{\m\s}(F^c)_{\s\n}
\end{equation}
The above equation leads to
\begin{equation} \label{eq:D^2->F}
\DC^2 (F^b)_{\m\n}(F^c)(F^d)(F^e)g(b,c,d,e)=- (F^a)_{\m\s}(F^b)_{\s\n}(F^c)(F^d)(F^e)
g([a,b],c,d,e)
\end{equation}
Thus in order to study corrections to the symmetrized trace prescription one has to consistently take into account derivative contributions. In particular the above relations clearly show that for the case of two covariant derivatives one has to consider not only the antisymmetrized products $\nabla_{[\m }\nabla_{\n]}$, as it was already noticed \cite{BRS,STT}, but also the  symmetrized ones $\nabla_{(\m }\nabla_{\n)}$. 

Using (\ref{eq:D^2->F}), we can rewrite the last two lines of the two derivative result in (\ref{eq:p^2finale}) as
\begin{eqnarray}
\label{eq:D^2F->FF}
&& -\frac{1}{20}\left(\frac{\pi^2}{12}\frac{1}{M^6}\right)\left(2 (F^5)_{a_1a_2a_3a_4a_5} + 2 (F^5)_{a_1a_2a_4a_3a_5} +4 (F^5)_{a_1a_4a_5a_2a_3} \right. \nonumber \\
&&\left.  +2 (F^2)_{a_3a_4}(F^3)_{a_1a_2a_5} + (F^2)_{a_2a_5}(F^3)_{a_1a_3a_4} \right)~g(a_1,a_2,a_3,a_4,a_5)
\end{eqnarray}
These terms can be combined with the $F^5$ contributions we have obtained from the five-point function in (\ref{final5point}). Making use of  the identities (\ref{eq:identities}) the sum of (\ref{eq:D^2F->FF}) and of (\ref{final5point}) can be written as
\begin{eqnarray}
&& \frac{1}{20}\left(\frac{\pi^2}{12}\frac{1}{M^6}\right)\left(-\frac{1}{2} (F^5)_{a_1a_2a_4a_3a_5}-\frac{1}{2}(F^5)_{a_1a_4a_2a_5a_3} \right. \nonumber \\
&& \left. ~~~~~~~~~~~~~~+ (F^5)_{a_1a_3a_2a_5a_4} \right)~g(a_1,a_2,a_3,a_4,a_5) \nonumber \\
&& =\frac{1}{20}\left(\frac{\pi^2}{12}\frac{1}{M^6}\right)\Tra \left(
-\frac{1}{2}F_{\m\n}F_{\n\rho}F_{\s\t}F_{\rho\s}F_{\t\m} 
 \right. \nonumber \\
&& \left.-\frac{1}{2}F_{\m\n}F_{\rho\s}F_{\t\m}F_{\n\rho}F_{\s\t} 
+ F_{\m\n}F_{\rho\s}F_{\n\rho}F_{\t\m}F_{\s\t} \right)
\label{F^5}
\end{eqnarray}
Thus our final result  to the $M^{-6}$ order is:
\begin{eqnarray}
\label{eq:totale}
\G_{tot}&=& \frac{1}{20}\left(\frac{\pi^2}{12}\frac{1}{M^6}\right) \Tra \Big\{
(\DC F_{\m\n} \DC F_{\n\rho}F_{\rho\s}F_{\s\m}
+\DC F_{\m\n} \DC F_{\rho\s}F_{\n\rho}F_{\s\m}+ F_{\m\n}\DC F_{\rho\s} \DC F_{\n\rho}F_{\s\m})
 \nonumber \\
&&\left. -\frac{1}{4}(\DC F_{\m\n} \DC F_{\m\n}F_{\rho\s}F_{\rho\s}
+F_{\m\n}\DC F_{\m\n}\DC F_{\rho\s}F_{\rho\s}
+\DC F_{\m\n}\DC F_{\rho\s}F_{\m\n}F_{\rho\s})\right\} \nonumber \\
&& +\frac{1}{20}\left(\frac{\pi^2}{12}\frac{1}{M^6}\right)\Tra \left(
 -\frac{1}{2} F_{\m\n}F_{\n\rho}F_{\s\t}F_{\rho\s}F_{\t\m} 
 \right. \nonumber \\
&& \left.~~~~~~~~~~~~~~-\frac{1}{2} F_{\m\n}F_{\rho\s}F_{\t\m}F_{\n\rho}F_{\s\t} 
+ F_{\m\n}F_{\rho\s}F_{\n\rho}F_{\t\m}F_{\s\t} \right)
\end{eqnarray}
Up to an overall numerical factor, the first two lines in (\ref{eq:totale}) reproduce the corresponding result in ref. \cite{kitazawa}, formula (3.3), obtained from an open superstring scattering amplitude. So, our first result is that, at the level of the four-point function, the supersymmetric Yang-Mills effective action exactly reproduces the structure of the non-abelian Born-Infeld theory, including the first derivative corrections. The terms $F^5$, which were also computed in \cite{kitazawa}, are more difficult to compare with ours: this is essentially due to the fact that the result quoted in \cite{kitazawa} is not written in a canonical form and moreover it requires additional symmetrizations which we are not clear how to interpret unambiguously.

We would like to conclude summarizing our results.\\
We have considered the ${\cal N}=4$ supersymmetric Yang-Mills theory and computed at one-loop the four- and five-point functions with external vector fields. From the superfield result we have extracted the part of the bosonic components which contain the field strengths $F_{\m\n}$. These non-local one-loop contributions have been expanded in a low energy approximation and expressed as a sum of an infinite series of local terms. We have argued that these local expressions reproduce contributions to the non-abelian Born-Infeld action, if supersymmetry has to determine its structure.
We have explicitly computed the leading contributions from the four- and the five-point functions and the subleading terms of the four-point function. These latter ones contain two derivatives acting on the four $F_{\m\n}$. From the four-point function calculation one simply obtains ordinary derivatives, thus the result is not in a gauge invariant form. We have checked that the correct covariantization of the result is obtained via contributions from the five- and the six-point functions with one and two connections respectively.

We have confirmed that beyond the $F^4$ order the non-abelian Born-Infeld action contains terms which are not included in the symmetrized trace prescription. In order to compute corrections one has to consider terms with derivatives, and in particular we have shown that one cannot disregard terms with symmetrized derivatives. Both the antisymmetrized products $\nabla_{[\m }\nabla_{\n]}$ and the  symmetrized ones $\nabla_{(\m }\nabla_{\n)}$ are on equal footing. 

The part of the six-point function needed in order to check the gauge invariance of our result has been computationally rather difficult, but clearly not impossible.
Taking advantage of the superfield approach and of the background field method 
the determination of the full six-point function seems at hand.

\vspace{0.5cm}

\noindent {\bf Acknowledgements}

\noindent The authors have
been partially supported by INFN, MURST, and the European 
Commission RTN program HPRN-CT-2000-00113 in which they are 
associated to the University of Torino. 
\vspace{1cm}
\appendix
\section{Notation and conventions}
\label{notations}
Our notations and  conventions are the ones in \cite{superspace}. We use a metric 
\begin{equation}
\e_{\m\n} = (-,+,\cdots,+)
\end{equation}
and raise and lower
spinor indices  with
\begin{eqnarray}
C_{\a\b} = \left(\begin{array}{cc} 0 & -i \\ i & 0 \end{array} \right)  &;& 
C^{\a\b} = \left(\begin{array}{cc} 0 & i \\ -i & 0 \end{array} \right) \nonumber \\
C_{\ad\bd} = C_{\a\b}  &;& 
C^{\ad\bd} = C^{\a\b} 
\end{eqnarray}
The $\s$ matrices are defined in terms of the Pauli matrices
\begin{eqnarray}
(\s_{\m})_{\a\ad} &\equiv& (\1,\vec{\s}) \nonumber \\
(\sb_{\m})^{\ad\a} &\equiv& C^{\ad\bd} C^{\a\b}(\s_{\m})_{\b\bd}
                 =(-\1,\vec{\s}) \nonumber \\
(\s_{\m\n})_{\a}^{~\b} & \equiv & 
     \frac{1}{4}(\s_{\m}\sb_{\n}-\s_{\n}\sb_{\m})_{\a}^{~\b} \nonumber \\
(\sb_{\m\n})^{\ad}_{~\bd} & \equiv & 
     \frac{1}{4}(\sb_{\m}\s_{\n}-\sb_{\n}\s_{\m})^{\ad}_{~\bd}
\label{sigmamatr}
\end{eqnarray}
Vector indices are transformed into spinor notation by
\begin{eqnarray}
V_{\a\ad}=(\s^{\m})_{\a\ad}V_{\m} &\qquad\qquad& 
V_{\m} = \frac{1}{2} (\sb_{\m})^{\ad\a} V_{\a\ad}
\end{eqnarray}
For an antisymmetric tensor of rank two we have
\begin{eqnarray} 
F_{\a\b\ad\bd} & = & 2f_{\a\b} C_{\ad\bd} +  2\bar f_{\ad\bd} C_{\a\b}
\end{eqnarray} 
where $f_{\a\b}$ and $\bar f_{\ad\bd}$ are symmetric bispinors. The above relations can be inverted
\begin{eqnarray} \label{eq:f->F}
f_{\a\b} &=& \frac{1}{2} (\s_{\m\n})_{\a\b} F^{\m\n} \nonumber\\
\bar f_{\ad\bd} &=& -\frac{1}{2} (\sb_{\m\n})_{\ad\bd} F^{\m\n} 
\end{eqnarray} 
Therefore an antisymmetric tensor is always expressible in terms of two symmetric bispinors.
\begin{equation} \label{eq:F->f}
F^{\m\n} =
   (\s_{\m\n})_{\a\b} f^{\a\b} - (\sb_{\m\n})_{\ad\bd} \bar f^{\ad\bd} 
\end{equation}
In terms of the  Levi-Civita tensor
$\epsilon_{\m\n\rho\s} \Rightarrow  \epsilon_{0123}=1$ 
we define the Hodge-dual of $F_{\m\n}$ as
\begin{equation}
(\star F)_{\m\n}=
  \wt{F}_{\m\n}\equiv \frac{1}{2} \epsilon_{\m\n}^{~~~\rho\s}F_{\rho\s}
\end{equation}
Using the relation
\begin{eqnarray}
\epsilon^{\m\n\rho\s}\epsilon_{\a\b\gamma\d} &=& 
  - \d_{\a}^{[\m}\d_{\b}^{\n}\d_{\gamma}^{\rho}\d_{\d}^{\s]}
\end{eqnarray}
one obtains the general formula
\begin{eqnarray}
  \label{eq:2tilde1}
  (\wt{F}_a)_{\a\b}(\wt{F}_b)^{\m\n} &= & \frac{1}{4} \epsilon^{\m\n\rho\s}\epsilon_{\a\b\gamma\d}(F_a)^{\gamma\d}(F_b)_{\rho\s} = \nonumber \\
&=& \frac{1}{2} (F^2)_{ab}[\d_{\b}^{\m}\d_{\a}^{\n} - \d_{\a}^{\m}\d_{\b}^{\n}] - [\d_{\a}^{\m}(F_a)^{\n\rho}(F_b)_{\rho\b} - 
\d_{\b}^{\m}(F_a)^{\n\rho}(F_b)_{\rho\a}] + \nonumber \\
& & + [\d_{\a}^{\n}(F_a)^{\m\rho}(F_b)_{\rho\b} - 
\d_{\b}^{\n}(F_a)^{\m\rho}(F_b)_{\rho\a}] - (F_a)^{\m\n}(F_b)_{\a\b}
\end{eqnarray}
and also
\begin{eqnarray}\label{eq:2tilde2}
(\wt{F}_a F_b \wt{F}_c)_{\m\n} & = & - (F_a F_b F_c)_{\m\n} \nonumber\\
(\wt{F}_a)_{\m\rho}(\wt{F}_b)^{\rho\n} & = & \frac{1}{2}(F^2)_{ab} \d_{\m}^{\n}
+ (F_a)^{\n\rho}(F_b)_{\rho\m}
\end{eqnarray}

\section{Further identities involving the $\s_{\m\n}$ matrices}
\label{s-algebra}
The basic relations we use are
\begin{eqnarray} \label{eq:base}
-\s_{\m} \sb_{\n} \s_{\rho} &=& (\eta_{\m\rho} \s_{\n} - \eta_{\n\rho} \s_{\m} - \eta_{\m\n} \s_{\rho}) - i \epsilon_{\m\n\rho\t}\s^{\t}  \\
- \sb_{\m} \s_{\n} \sb_{\rho} &=& (\eta_{\m\rho} \sb_{\n} - \eta_{\n\rho} \sb_{\m} - \eta_{\m\n} \sb_{\rho}) + i \epsilon_{\m\n\rho\t}\sb^{\t}
\end{eqnarray}
These allow to obtain the following results for the trace of two
$\s_{\m\n}$ matrices
\begin{eqnarray}
\label{eq:B4}
&& \Tr(\s_{\m\n}\s_{\rho\s}) = 
 +\frac{1}{4}\left[\eta_{\m\s}\eta_{\n\rho} + \frac{i}{2}\epsilon_{\m\n\rho\s} \right] \nonumber \\
&&  ~~~~~~~~~~~~~~~~~~~~~~~~+\textrm{completely antisymmetrized in }(\m,\n)(\rho,\s)  
\end{eqnarray}
and similarly
\begin{eqnarray}
\label{eq:B4b}
&& \Tr(\sb_{\m\n}\sb_{\rho\s}) = 
 +\frac{1}{4}\left[\eta_{\m\s}\eta_{\n\rho} - \frac{i}{2}\epsilon_{\m\n\rho\s} \right] \nonumber \\
&&  ~~~~~~~~~~~~~~~~~~~~~~~~+\textrm{completely antisymmetrized in }(\m,\n)(\rho,\s)  
\end{eqnarray}
For the trace of three $\s_{\m\n}$ matrices we obtain
\begin{eqnarray}
  \label{eq:B6}
 &&\Tr(\s_{\m\n}\s_{\rho\s}\s_{\t\epsilon}) = \nonumber \\
&&~~~~~= \left[ \frac{1}{8}\eta_{\m\rho}\eta_{\s\epsilon}\eta_{\n\t} + \frac{i}{16}(\eta_{\m\rho}\epsilon_{\s\t\epsilon\n} +\eta_{\s\t}\epsilon_{\m\n\rho\epsilon}) + \frac{1}{32} \epsilon_{\m\n\rho\eta} \epsilon_{\s\t\epsilon}^{~~~\eta} \right] \nonumber \\
& & ~~~~~~~~~~~~+ \textrm{completely antisymmetrized in}(\m,\n) (\rho,\s)(\t,\eta)
\end{eqnarray}
and similarly
\begin{eqnarray}
  \label{eq:Bb6}
&& \Tr(\sb_{\m\n}\sb_{\rho\s}\sb_{\t\epsilon}) = \nonumber \\
&&~~~~~= \left[ \frac{1}{8}\eta_{\m\rho}\eta_{\s\epsilon}\eta_{\n\t} - \frac{i}{16}(\eta_{\m\rho}\epsilon_{\s\t\epsilon\n} +\eta_{\s\t}\epsilon_{\m\n\rho\epsilon}) + \frac{1}{32} \epsilon_{\m\n\rho\eta} \epsilon_{\s\t\epsilon}^{~~~\eta} \right] \nonumber \\
& & ~~~~~~~~~~~~+ \textrm{completely antisymmetrized in}(\m,\n) (\rho,\s)(\t,\eta)
\end{eqnarray}
\section{Special identities for the $F_{\m\n}$ field strengths}
\label{identities}
In this Appendix we study structures with five $F_{\m\n}$ field strengths. We can contract the $\m\n$ indices in different ways, moreover we can place the various $F_{\m\n}$'s in different positions with respect to the gauge group trace.
For example we can write
\begin{eqnarray}
\Tra (F_{\m\n}F_{\n\rho}F_{\rho\s}F_{\s\t}F_{\t\m}) &=& (F^5)_{a_1a_2a_3a_4a_5}g(a_1,a_2,a_3,a_4,a_5) \nonumber \\
\Tra (F_{\m\n}F_{\n\rho}F_{\rho\m}F_{\s\t}F_{\t\s}) &=& (F^3)_{a_1a_2a_3}(F^2)_{a_4a_5}g(a_1,a_2,a_3,a_4,a_5)
\end{eqnarray}
Clearly not all the possible structures are independent, since we can freely use the ciclicity and inversion property of the trace. 
\begin{figure}[htb] \begin{center}
\mbox{\epsfig{figure={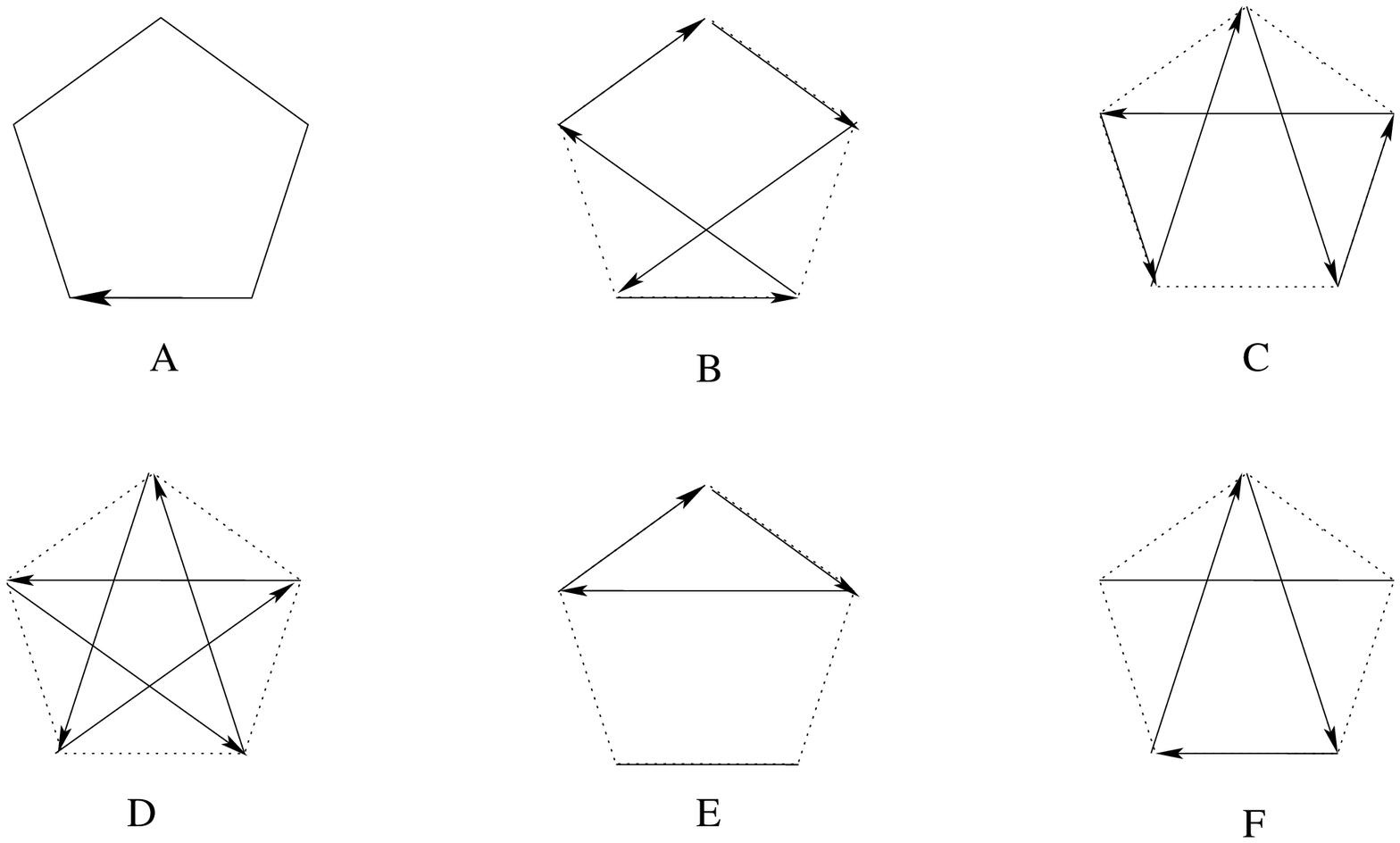},
      width=0.77 \linewidth}}
\end{center}
\begin{center}
{\small {Figure 4: $F^5$ structures }}
\end{center}
\end{figure}
It is easy to realize that among the $F^5$ structures only four are independent: we can list them as follows 
\begin{eqnarray}
  \label{eq:F^5}
A:~~\Tra (F_{\m\n}F_{\n\rho}F_{\rho\s}F_{\s\t}F_{\t\m}) &=& (F^5)_{a_1a_2a_3a_4a_5}g(a_1,a_2,a_3,a_4,a_5) \nonumber\\
B:~~\Tra (F_{\m\n}F_{\n\rho}F_{\rho\s}F_{\t\m}F_{\s\t}) &=&(F^5)_{a_1a_2a_3a_5a_4}g(a_1,a_2,a_3,a_4,a_5)  \nonumber\\
C:~~\Tra (F_{\m\n}F_{\rho\s}F_{\n\rho}F_{\t\m}F_{\s\t}) &=&(F^5)_{a_1a_3a_2a_5a_4}g(a_1,a_2,a_3,a_4,a_5) \nonumber\\
D:~~\Tra (F_{\m\n}F_{\rho\s}F_{\t\m}F_{\n\rho}F_{\s\t}) &=& (F^5)_{a_1a_4a_2a_5a_3}g(a_1,a_2,a_3,a_4,a_5) 
\end{eqnarray}
In the same way one finds that there exist only two independent structures $F^2 F^3$ 
\begin{eqnarray}
  \label{eq:F^2F^3}
E:~~\Tra (F_{\m\n}F_{\n\rho}F_{\rho\m}F_{\s\t}F_{\s\t}) &=& (F^3)_{a_1a_2a_3}(F^2)_{a_4a_5}g(a_1,a_2,a_3,a_4,a_5) \nonumber\\
F:~~\Tra (F_{\m\n}F_{\s\t}F_{\n\rho}F_{\rho\mu}F_{\s\t}) &=&(F^3)_{a_1a_3a_4}(F^2)_{a_2a_5}g(a_1,a_2,a_3,a_4,a_5)
\end{eqnarray}
The terms in (\ref{eq:F^5}) and (\ref{eq:F^2F^3}) can be represented graphically as shown in Fig.4 and explained in \cite{STT}. 
Every vertex corresponds to a $F_{\m\n}$ field strength and the trace operation has to be taken following the vertices of the pentagon clockwise. The arrows denote how the $\m$$\n$ indices are contracted.
\begin{figure}[htb] \begin{center}
\mbox{\epsfig{figure={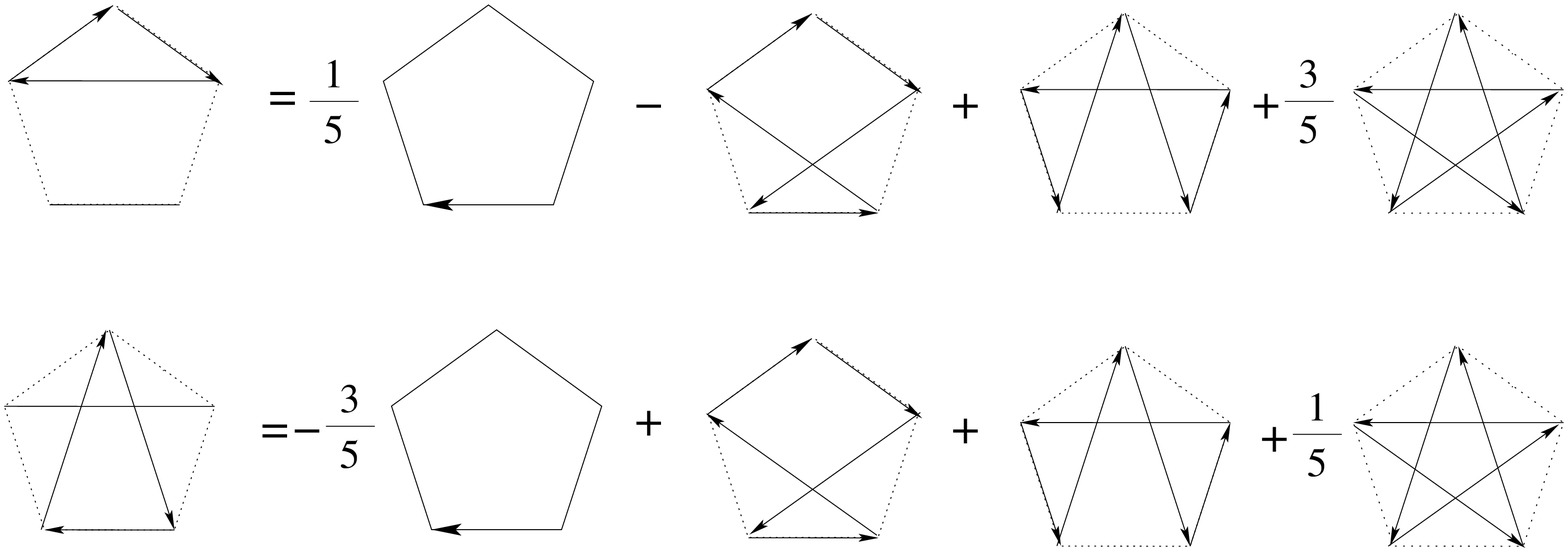},
      width=0.85 \linewidth}}
\end{center}
\begin{center}
{\small {Figure 5: Special identities}}
\end{center}
\end{figure}
A quite remarkable result is that the $F^2 F^3$ structures can be reexpressed in terms of the $F^5$ structures as follows:
\begin{equation}
\label{eq:identities}
F= \frac{1}{5}A -B + C +\frac{3}{5} D \qquad \qquad  \qquad 
E= -\frac{3}{5}A + B + C +\frac{1}{5} D  
\end{equation}
We can give a graphical representation of these relations as is shown in Fig.5.
So actually there are only $4$ independent structures containing $5$ $F_{\m\n}$.

\end{document}